\documentclass[aps,prd,twocolumn,groupedaddress,nofootinbib,longbibliography]{revtex4-2}
\usepackage{amsmath,amssymb}
\usepackage{graphicx}
\usepackage{bm}
\usepackage{hyperref}
\usepackage{xcolor}
\usepackage{multirow}
\usepackage{dcolumn}
\usepackage[utf8]{inputenc}

\newcommand{\mub}{\mu_B}
\newcommand{\muq}{\mu_Q}
\newcommand{\mus}{\mu_S}
\newcommand{\muhat}{\hat{\mu}_B}
\newcommand{\Tpc}{T_{\mathrm{pc}}}
\newcommand{\Tdec}{T_{\mathrm{dec}}}
\newcommand{\Dls}{\Delta_{l,s}}
\newcommand{\Ploop}{\Phi}
\newcommand{\Ploopbar}{\bar{\Phi}}
\newcommand{\chiB}[1]{\chi_{#1}^{B}}
\newcommand{\chiQ}[1]{\chi_{#1}^{Q}}
\newcommand{\chiS}[1]{\chi_{#1}^{S}}

\newcommand{\chioff}[2]{\chi_{#1}^{#2}}
\newcommand{\SB}{\mathrm{SB}}

\newcommand{\Nc}{N_c}

\begin{document}
	
	
	\title{Fluctuations and correlations of conserved charges in the Polyakov chiral SU(3) quark mean field model}
	
	\author{Dhananjay Singh}
	\email{snaks16aug@gmail.com}
	\affiliation{Department of Physics, Dr. B. R. Ambedkar National Institute of Technology, Jalandhar - 144008, India}
	
	\author{Arvind Kumar}
	\email{kumara@nitj.ac.in}
	\affiliation{Department of Physics, Dr. B. R. Ambedkar National Institute of Technology, Jalandhar - 144008, India}
	
	\date{\today}
	
	\begin{abstract}
		We compute generalized susceptibilities of conserved charges in the Polyakov chiral SU(3) quark mean field (PCQMF) model with the fermion vacuum term. At $\mu_B = 0$ MeV, the calculation covers the diagonal $\chi_n^{B,Q,S}$ through eighth order and all twelve independent fourth-order off-diagonal correlators. Extending to finite $\mu_B$ at $\mu_Q = \mu_S = 0$, we compute $\chi_n^B$ through eighth order, $\chi_n^{Q,S}$ through fourth order, the second-order off-diagonals, all twelve fourth-order off-diagonal correlators, and the odd-order baryon susceptibilities $\chi_1^B$, $\chi_3^B$, $\chi_5^B$. The calculation includes the vacuum term (vac=1) and is repeated for an independently refitted no-sea variant (vac=0). At $\mu_B = 0$ MeV, the chiral pseudocritical temperature is $T_{\mathrm{pc}} = 170.5$ MeV (vac=1) and $166.4$ MeV (vac=0), while the Polyakov-loop deconfinement temperature is $T_{\mathrm{dec}} = 144.4$ MeV (vac=1) and $146.6$ MeV (vac=0). In vac=1, the derivative $-d\Delta_{l,s}/dT$ of the subtracted chiral condensate develops an inflection near $T_{\mathrm{dec}}$. Higher derivative orders resolve the chiral-deconfinement splitting as twin maxima in $\chi_4^B$ and $\chi_6^Q$, twin minima in $\chi_8^B$ and $\chi_8^Q$, and multiple zero crossings in $\chi_6^B$. Among the fourth-order off-diagonal correlators, vac=1 amplitudes exceed vac=0 in the BQ channel across the chiral crossover. The BS, QS, and mixed BQS components peak near the strange-melting temperature, where vac=0 dominates. Along $T_{\mathrm{pc}}(\mu_B)$, the kurtosis ratio $R_{42}^B \equiv \chi_4^B/\chi_2^B$ of vac=1 crosses zero at $\mu_B/T_{\mathrm{pc}} \approx 2.15$, while vac=0 stays positive across the full range. The higher-order ratios $R_{51}^B \equiv \chi_5^B/\chi_1^B$ and $R_{62}^B \equiv \chi_6^B/\chi_2^B$ start negative in vac=1 and grow more negative as $\mu_B$ increases.
	\end{abstract}
	
	\maketitle
	
	\section{Introduction}
	\label{sec:intro}
	
	Quantum chromodynamics (QCD) at finite temperature $T$ and baryon chemical potential $\mub$ describes matter under conditions realized in the early universe and in relativistic heavy-ion collisions. At high temperature, asymptotic freedom implies deconfinement into a quark-gluon plasma, and chiral symmetry is expected to be approximately restored. Lattice simulations at $\mub = 0$ MeV have established that the two transitions (chiral and deconfinement) coincide as a smooth crossover \cite{Aoki:2006we,Bazavov:2019PLB,Borsanyi:2010WB}. The pseudocritical temperature at $\mu_B = 0$ is $\Tpc = 158.0 \pm 0.6$ MeV \cite{Borsanyi:2020WB_PRL}. At finite $\mub$, effective models and universality arguments suggest a first-order phase boundary that terminates at a critical endpoint (CEP) \cite{Halasz:1998,Stephanov:1998,Stephanov:2006}. The fermion sign problem prevents direct lattice simulations at finite $\mub$. The lattice equation of state (EoS) now extends to $\mub/T \approx 3.5$ \cite{Abuali:2025WB}, and recent Pad\'{e} and Lee-Yang analyses find a CEP unlikely below that reach \cite{Clarke:2025,Basar:2024}. On the experimental side, the RHIC Beam Energy Scan program has measured net-proton cumulants at $\sqrt{s_{NN}} = 7.7$-200 GeV, where a non-monotonic $\kappa\sigma^2$ was first seen in BES-I \cite{Adam:2021} and confirmed in BES-II \cite{STAR:2025_BES2}, which also reported the first fifth- and sixth-order results \cite{STAR:2025_C5C6}. Fixed-target measurements extend the reach to $\sqrt{s_{NN}} = 3$ GeV \cite{STAR:2023_C4C2}. Upcoming programs at FAIR (CBM) and NICA will extend the reach to higher $\mub$ (lower $\sqrt{s_{NN}}$) \cite{Bzdak:2020,Pandav:2022}. Recent reviews are given in \cite{Luo:2017,Koch:2025_review}. These data call for theoretical predictions for the full set of generalized baryon-charge-strangeness (BQS) susceptibilities, both diagonal and off-diagonal. Cumulants of net-proton, net-kaon, and net-charge distributions measured in heavy-ion collisions serve as experimental proxies for specific components of this set \cite{Koch:2008,Kitazawa:2012}.
	
	The generalized susceptibilities $\chi_{ijk}^{BQS}$ are derivatives of the scaled pressure with respect to $\hat{\mu}_{B,Q,S} = \mu_{B,Q,S}/T$. They can be computed in both lattice QCD and effective models, and correspond directly to the cumulants of event-by-event net-particle distributions \cite{Koch:2005,Ejiri:2006,Koch:2008,Hatta:2003}. At second order, the diagonal $\chi_2^{B,Q,S}$ have been obtained on the lattice at $\mub = 0$ MeV \cite{Bazavov:2012,Borsanyi:2012,Bellwied:2015WB,Bollweg:2021HotQCD}, at finite $\mub$ via Taylor expansion \cite{Karsch:2011,Bazavov:2020_skewness_kurtosis}, and across the full $(T, \mu_B, \mu_Q, \mu_S)$ plane \cite{Abuali:2025WB}. In the Polyakov Nambu Jona Lasinio (PNJL) model \cite{Fukushima:2004,Ratti:2006}, diagonal susceptibilities were computed with multi-quark interactions \cite{Ghosh:2006,Bhattacharyya:2010}, where eight-quark terms improved agreement with lattice data below $\Tpc$. Later PNJL studies covered net-baryon higher moments \cite{Sarkar:2024} and net-strangeness fluctuations \cite{Sarkar:2026} at BES energies. Off-diagonal correlators and the $C_{BS}$ ratio at finite $\mub$ have also been computed \cite{Shao:2018,Ferreira:2018}. In the Polyakov quark meson (PQM) model \cite{Schaefer:2007_Polyakov,Schaefer:2010_thermo}, diagonal susceptibilities and their ratios were computed with and without the fermion vacuum term \cite{Schaefer:2012,Chatterjee:2012}. In the hadron resonance gas (HRG) model, repulsive mean-field interactions are essential for reproducing higher-order susceptibilities near $\Tpc$ \cite{Pal:2021}. 
	
	The second-order off-diagonal correlators $\chi_{11}^{BQ,BS,QS}$ have received much less attention. They probe correlated fluctuations between conserved charges and are sensitive to the quantum numbers of the dominant carriers. For instance, the baryon-strangeness correlation coefficient $C_{BS}$ distinguishes whether strangeness is carried by hadrons or by quarks \cite{Koch:2005}. It has been computed in the PNJL model and on the lattice along the pseudocritical line \cite{Bollweg:2024_strangeness,Goswami:2025}, and baryon-charge correlations have been examined in nuclear matter \cite{Yang:2025}. A persistent deficiency of these mean-field chiral models is the underestimation of $\chi_2^Q$ below $\Tpc$ \cite{Chatterjee:2012,Bhattacharyya:2010}, since pions in mean-field PCQMF are frozen at their classical value and contribute no fluctuations. 
	
	Higher-order susceptibilities are directly sensitive to critical dynamics. The kurtosis $\chi_4^B/\chi_2^B$ develops non-monotonic structure near the chiral crossover \cite{Stephanov:2009,Stephanov:2011}, and $\chi_6^B$ changes sign near $\Tpc$, a feature linked to $O(4)$ universality \cite{Ejiri:2006,Friman:2011}. Continuum-extrapolated lattice results for $\chi_6^B$ and $\chi_8^B$ now exist \cite{Borsanyi:2024_chi68}, and $\chi_6$ through $\chi_{10}$ have been measured on finite-volume lattices \cite{Adam:2025_chi610}. Lattice skewness and kurtosis ratios at finite $\mub$ have been compared with STAR BES-I data \cite{Bazavov:2020_skewness_kurtosis}. More recently, higher-order cumulants have been evaluated along the pseudocritical line \cite{Goswami:2025}. In effective models, the scope of existing calculations varies considerably. In the PQM model, all twelve fourth-order off-diagonal correlators and diagonal susceptibilities through sixth order were computed with the vacuum term at $\mub = 0$ MeV \cite{Chatterjee:2012}. A separate calculation extended baryon-number ratios to twelfth order at finite $\mub$ along a single quark chemical potential \cite{Schaefer:2012}. In the NJL model, diagonal BQS susceptibilities were computed through fourth order \cite{Fan:2017} and baryon-number susceptibilities through eighth order \cite{Fan:2019}. Beyond mean field, functional renormalization group (fRG) calculations resum mesonic and fermionic quantum fluctuations and have produced baryon fluctuations through tenth order \cite{Fu:2021_hyperorder,Fu:2020,Fu:2025_ripples}. The same approach has been applied to strangeness neutrality and baryon-strangeness correlations \cite{Fu:2020_strangeness_phasediag,Fu:2026}. Holographic QCD has produced baryon susceptibilities through twelfth order \cite{Li:2023_holographic}. No existing effective-model study provides all twelve fourth-order off-diagonal BQS correlators together with diagonal susceptibilities through eighth order at $\mu_B = 0$.
	
	The vacuum (Dirac-sea) term, the zero-temperature one-loop quark contribution to the thermodynamic potential, significantly affects the crossover. In the two-flavor PQM model, the vacuum term restores second-order $O(4)$ universality in the chiral limit \cite{Skokov:2010_vacuum}. In the $(2+1)$-flavor extension, it smooths the crossover and produces oscillatory sign structure in $\chi_6^{B,Q,S}$ \cite{Chatterjee:2012,Schaefer:2012}, and improves agreement with lattice thermodynamics \cite{Chatterjee:2012_thermo}. It also shifts the CEP to larger $\mub$ and lower $T$ \cite{Gupta:2012,Schaefer:2012}, though the magnitude of the shift depends on the renormalization scheme \cite{Rai:2024}. In the Polyakov chiral SU(3) quark mean field (PCQMF) model, the vacuum term has only been used for transport coefficients \cite{Singh:2025} and anisotropic quark matter \cite{Singh:2026}, while all existing thermodynamic and susceptibility studies \cite{Kumari:2021,Chahal:2022,Chahal:2023} omitted it. Its effect on the full set of generalized BQS susceptibilities has not been studied.
	
	The PCQMF model includes the nonstrange and strange scalar fields $\sigma$ and $\zeta$ as order parameters for chiral symmetry breaking, and the Polyakov loop $\Phi$ as an order parameter for deconfinement. Vector mesons $\omega$, $\rho$, $\phi$ provide short-range repulsion, and a dilaton field $\chi$ implements the QCD trace anomaly. An isovector scalar $\delta$ is included for isospin-asymmetric matter. The only existing susceptibility calculation in this model \cite{Chahal:2022} omitted the vacuum term and computed only diagonal susceptibilities, through sixth order for charge and strangeness and eighth order for baryon. No off-diagonal correlators were reported. We report the first computation of the full set of generalized BQS susceptibilities in the PCQMF model with the fermion vacuum term. At $\mub = 0$ MeV, the calculation covers all diagonal $\chi_n^{B,Q,S}$ for $n = 2, 4, 6, 8$ and all twelve independent fourth-order off-diagonal correlators, namely the nine two-charge correlators $\chi_{ij}^{BQ,BS,QS}$ with $i+j = 4$ and the three mixed correlators $\chi_{ijk}^{BQS}$ with $i+j+k = 4$. The extension to finite $\mub$ up to 500 MeV at $\muq = \mus = 0$ covers $\chi_2^{B,Q,S}$, $\chi_4^{B,Q,S}$, $\chi_6^B$, $\chi_8^B$, the second-order off-diagonals $\chi_{11}^{BQ,BS,QS}$, and the odd-order baryon susceptibilities $\chi_1^B$, $\chi_3^B$, $\chi_5^B$. The calculation is repeated with an independently refitted no-sea (vac=0) parameter set for comparison. We compare with continuum-extrapolated lattice data \cite{Bellwied:2015WB,Borsanyi:2024_chi68,Bollweg:2021HotQCD,Abuali:2025WB} where benchmarks exist.
	
	The article is organized as follows. Section \ref{sec:model} introduces the chiral SU(3) Lagrangian, the PCQMF grand potential with the vacuum term, the coupled gap equations, and the susceptibility definitions. Section \ref{sec:results_mub0} presents results at $\mub = 0$ MeV, including order parameters, bulk thermodynamics, diagonal susceptibilities through eighth order, all twelve fourth-order off-diagonal correlators, and correlation ratios. 
	Section \ref{sec:results_finite_mub} extends the analysis to finite $\mub$, covering diagonal and off-diagonal susceptibilities at second and fourth order, diagonal baryon susceptibilities through eighth order, and cumulant ratios along $\Tpc(\mub)$. Section \ref{sec:discussion} discusses the vacuum-term effects and mean-field limitations. We conclude in Sec. \ref{sec:summary} with a summary and outlook. Stefan-Boltzmann limits for the quantities computed in this work are collected in Appendix \ref{app:SB}.
	
	\section{Model and formalism}
	\label{sec:model}
	
	\subsection{Chiral SU(3) Lagrangian}
	\label{sec:lagrangian}
	
	We employ the Polyakov chiral SU(3) quark mean field (PCQMF) model. The chiral SU(3) Lagrangian \cite{Papazoglou:1998,Papazoglou:1999} uses a nonlinear realization of chiral symmetry. Baryons couple to scalar mesons $\sigma$, $\zeta$, $\delta$ and vector mesons $\omega$, $\rho$, $\phi$. The dilaton $\chi$ accounts for broken scale invariance. A later reformulation at the quark level \cite{Wang:2001,Wang:2003} replaced the baryons with constituent quarks, and the Polyakov-loop variables ($\Ploop$, $\Ploopbar$) \cite{Fukushima:2004,Ratti:2006} were added in Ref. \cite{Kumari:2021}. The fermion vacuum (Dirac-sea) term, first included in the PCQMF model in Ref. \cite{Singh:2025}, is retained throughout the present work.
	
	The Lagrangian comprises scalar and vector meson self-interactions, explicit chiral symmetry breaking, and a strange-quark mass term:
	\begin{equation}
		\mathcal{L}_{\mathrm{eff}}
		= \mathcal{L}_{\Sigma\Sigma}
		+ \mathcal{L}_{VV}
		+ \mathcal{L}_{SB}
		+ \mathcal{L}_{\Delta m}.
		\label{eq:L_eff}
	\end{equation}
	The scalar sector reads
	\begin{align}
		\mathcal{L}_{\Sigma\Sigma}
		&= -\tfrac{1}{2}k_0\chi^2(\sigma^2+\zeta^2+\delta^2)
		+ k_1(\sigma^2+\zeta^2+\delta^2)^2
		\nonumber\\
		&\quad + k_2\!\left(\frac{\sigma^4}{2}+\frac{\delta^4}{2}
		+ 3\sigma^2\delta^2 + \zeta^4\right)
		\nonumber\\
		&\quad + k_3\chi(\sigma^2-\delta^2)\zeta
		- k_4\chi^4
		- \frac{1}{4}\chi^4\ln\frac{\chi^4}{\chi_0^4}
		\nonumber\\
		&\quad + \frac{d}{3}\chi^4
		\ln\!\left[\frac{(\sigma^2-\delta^2)\zeta}{\sigma_0^2\zeta_0}
		\cdot\frac{\chi^3}{\chi_0^3}\right],
		\label{eq:L_SS}
	\end{align}
	where $\sigma_0$ and $\zeta_0$ are the vacuum expectation values of $\sigma$ and $\zeta$ and given by $\sigma_0 = -f_\pi$ and $\zeta_0 = (f_\pi - 2f_K)/\sqrt{2}$. The dilaton vacuum value $\chi_0$ is fitted to vacuum observables and listed in Table \ref{tab:params}. The pion and kaon decay constants are $f_\pi = 93$ MeV and $f_K = 115$ MeV. The logarithmic and $k_4$ terms together implement broken scale invariance and generate the trace anomaly \cite{Papazoglou:1999,Wang:2001}.
	
	The vector meson self-interaction reads
	\begin{align}
		\mathcal{L}_{VV}
		&= \frac{1}{2}\frac{\chi^2}{\chi_0^2}
		\bigl(m_\omega^2\omega^2 + m_\rho^2\rho^2 + m_\phi^2\phi^2\bigr)
		\nonumber\\
		&\quad + g_4(\omega^4 + 6\omega^2\rho^2 + \rho^4 + 2\phi^4),
		\label{eq:L_VV}
	\end{align}
	with $m_\omega^2 = m_\rho^2 = m_v^2/(1 - \tfrac{1}{2}\mu_v\sigma^2)$ and $m_\phi^2 = m_v^2/(1 - \mu_v\zeta^2)$, where $m_v$ and $\mu_v$ are chosen to reproduce $m_\omega = 783$ MeV and $m_\phi = 1020$ MeV \cite{Kumari:2021}. Explicit chiral symmetry breaking enters through
	\begin{equation}
		\mathcal{L}_{SB} = -\frac{\chi^2}{\chi_0^2}\bigl(h_x\sigma + h_y\zeta\bigr),
		\label{eq:L_SB}
	\end{equation}
	where $h_x = m_\pi^2 f_\pi$ and $h_y = \sqrt{2}\,m_K^2 f_K - m_\pi^2 f_\pi/\sqrt{2}$ are fixed by the physical pion and kaon masses. A strange-quark mass term
	\begin{equation}
		\mathcal{L}_{\Delta m} = -\Delta m_s\,\bar{\psi}\,S\,\psi,
		\label{eq:L_Dm}
	\end{equation}
	with $S = \mathrm{diag}(0,0,1)$ and $\Delta m_s = 29$ MeV, supplies a bare mass to the strange quark.
	
	The quark-meson couplings follow from the SU(3) flavor structure,
	\begin{equation}
		g_{\sigma u} = g_{\sigma d} = g_{\delta u} = -g_{\delta d} = \frac{g_s}{\sqrt{2}},
		\quad
		g_{\zeta s} = g_s,
	\end{equation}
	with all other scalar couplings equal to zero and $g_s = 4.76$ as the single independent coupling. The vector coupling is set to $g_v = 0$.
	
	Numerical values of the parameters for both variants are collected in Table \ref{tab:params}. The fit imposes physical $\pi$ and $K$ masses, the combined $\eta$-$\eta'$ mass, the scalar $\sigma$-meson mass, and the vacuum stationarity conditions. The procedure was established for this model in Refs. \cite{Wang:2001, Kumari:2021}. The PCAC relations fix $h_x$ and $h_y$ to common values in both variants. The resulting vacuum constituent quark masses are $m_u^* = m_d^* = 313$ MeV and $m_s^* = 490$ MeV.
	
	\begin{table*}[!htbp]
		\caption{PCQMF model parameters for the variant with the fermion vacuum term (vac=1) and without it (vac=0). The explicit symmetry breaking parameters $h_x$, $h_y$ are common to both variants, as they are fixed by the $\pi$ and $K$ mass constraints.}
		\label{tab:params}
		\begin{ruledtabular}
			\begin{tabular}{lccc}
				Parameter & Symbol & vac=1 & vac=0 \\
				\hline
				Scalar potential couplings
				& $k_0$    & $2.002\times 10^{-1}$ & $3.881$  \\
				& $k_1$    & $2.388$               & $1.855$  \\
				& $k_2$    & $-19.50$              & $-10.19$ \\
				& $k_3$    & $-4.733$              & $-4.443$ \\
				& $k_4$    & $-0.060$              & $-0.060$ \\
				& $d$      & $0.002$               & $0.182$  \\
				Dilaton scale (MeV)
				& $\chi_0$ & $327.8$               & $278.2$  \\
				Explicit breaking (MeV$^3$)
				& $h_x$    & $1.797\times 10^{6}$  & $1.797\times 10^{6}$ \\
				& $h_y$    & $3.874\times 10^{7}$  & $3.874\times 10^{7}$ \\
				UV cutoff (MeV)
				& $\Lambda$ & $600$                & \textemdash \\
				\hline
				Vector sector
				& $m_v$     & $673.6$ MeV               & $673.6$ MeV               \\
				& $\mu_v$   & $2.34$ fm$^{2}$           & $2.34$ fm$^{2}$           \\
				& $g_4$                   & $37.5$               & $37.5$               \\
				& $g_v$                   & $0$                  & $0$                  \\
				Strange-quark shift (MeV)
				& $\Delta m_s$            & $29$                 & $29$                 \\
				\hline
				Polyakov potential
				& $a_0$        & $1.81$  & $1.81$  \\
				& $a_1$        & $-2.47$ & $-2.47$ \\
				& $a_2$        & $15.2$  & $15.2$  \\
				& $b_3$        & $-1.75$ & $-1.75$ \\
				& $T_0$ (MeV)  & $200$   & $200$   \\
			\end{tabular}
		\end{ruledtabular}
	\end{table*}
	
	\subsection{Thermodynamic potential}
	\label{sec:grand_potential}
	
	In the mean-field approximation, the thermodynamic potential density reads
	\begin{equation}
		\Omega = \Omega_{\mathrm{mes}}
		+ \Omega_{\bar{q}q}^{\mathrm{vac}}
		+ \Omega_{\bar{q}q}^{\mathrm{th}}
		+ \mathcal{U}(\Ploop, \Ploopbar; T),
		\label{eq:Omega_total}
	\end{equation}
	with $\Omega_{\mathrm{mes}}$ the mesonic potential, $\Omega_{\bar{q}q}^{\mathrm{vac}}$ the vacuum contribution, $\Omega_{\bar{q}q}^{\mathrm{th}}$ the thermal quark-antiquark contribution, and $\mathcal{U}(\Ploop, \Ploopbar; T)$ the Polyakov-loop effective potential. The constituent quark masses entering $\Omega_{\bar{q}q}^{\mathrm{vac}}$ and $\Omega_{\bar{q}q}^{\mathrm{th}}$ are
	\begin{equation}
		m_i^* = -g_{\sigma i}\sigma - g_{\zeta i}\zeta - g_{\delta i}\delta + m_{i0},
		\label{eq:mstar}
	\end{equation}
	with $m_{u0} = m_{d0} = 0$ and $m_{s0} = \Delta m_s$. The quark chemical potentials are
	\begin{equation}
		\mu_i = B_i\mub + Q_i\muq + S_i\mus,
		\label{eq:mu_relation}
	\end{equation}
	with $B_i = 1/3$ for all flavors, $Q_u = 2/3$, $Q_d = Q_s = -1/3$, $S_s = -1$, and $S_u = S_d = 0$, giving
	\begin{align}
		\mu_u &= \tfrac{1}{3}\mub + \tfrac{2}{3}\muq,
		\nonumber \\
		\mu_d &= \tfrac{1}{3}\mub - \tfrac{1}{3}\muq,
		\nonumber \\
		\mu_s &= \tfrac{1}{3}\mub - \tfrac{1}{3}\muq - \mus.
		\label{eq:mu_flavor}
	\end{align}
	Coupling to the vector mean fields shifts the chemical potentials to
	\begin{equation}
		\mu_i^* = \mu_i - g_\omega^i \omega - g_\rho^i \rho - g_\phi^i \phi,
		\label{eq:mustar}
	\end{equation}
	with the vector couplings $g_\omega^u = g_\omega^d = g_v/\sqrt{2}$, $g_\rho^u = -g_\rho^d = g_v/\sqrt{2}$, and $g_\phi^s = g_v$. All other vector couplings are set to zero.
	
	The mesonic potential follows from Eq. (\ref{eq:L_eff}) as
	\begin{equation}
		\Omega_{\mathrm{mes}} = -\mathcal{L}_{\mathrm{eff}} - \mathcal{V}_{\mathrm{vac}},
		\label{eq:Omega_mes}
	\end{equation}
	where $\mathcal{V}_{\mathrm{vac}}$ is the additive constant enforcing $\Omega = 0$ in vacuum.
	
	The thermal quark-antiquark contribution in Eq. (\ref{eq:Omega_total}), dressed by the Polyakov-loop variables, reads
	\begin{equation}
		\Omega_{\bar{q}q}^{\mathrm{th}} = -\sum_{i=u,d,s}\frac{\gamma_i T}{(2\pi)^3}\int_0^{\infty} d^3 k \bigl\{\ln\mathcal{N}_i^- + \ln\mathcal{N}_i^+\bigr\},
		\label{eq:Omega_th}
	\end{equation}
	with spin degeneracy $\gamma_i = 2$ and
	\begin{align}
		\mathcal{N}_i^- &= 1 + e^{-3 E_-} + 3\Ploop\,e^{-E_-} + 3\Ploopbar\,e^{-2 E_-}, \label{eq:N_minus} \\[4pt]
		\mathcal{N}_i^+ &= 1 + e^{-3 E_+} + 3\Ploopbar\,e^{-E_+} + 3\Ploop\,e^{-2 E_+}, \label{eq:N_plus}
	\end{align}
	where $E_\pm = (E_i^*(k) \pm \mu_i^*)/T$ and $E_i^*(k) = \sqrt{k^2 + m_i^{*2}}$ is the effective single-particle energy.
	
	The Polyakov-loop effective potential in Eq. (\ref{eq:Omega_total}) is taken in the logarithmic form \cite{Roessner:2007}
	\begin{align}
		\frac{\mathcal{U}(\Ploop, \Ploopbar; T)}{T^4}
		&= -\frac{a(T)}{2}\Ploop\Ploopbar
		+ b(T)\ln\!\bigl[1 - 6\Ploop\Ploopbar
		\nonumber \\
		&\quad + 4(\Ploop^3 + \Ploopbar^3)
		- 3(\Ploop\Ploopbar)^2\bigr],
		\label{eq:U_poly}
	\end{align}
	with $a(T) = a_0 + a_1(T_0/T) + a_2(T_0/T)^2$ and $b(T) = b_3(T_0/T)^3$. The logarithm argument is proportional to the $\mathrm{SU}(3)$ Haar measure, restricting $\Ploop$ and $\Ploopbar$ to $|\Ploop|, |\Ploopbar| \leq 1$. The parameters $a_0$, $a_1$, $a_2$, $b_3$ are fitted to pure-gauge lattice thermodynamics \cite{Roessner:2007}. To account for the backreaction of dynamical quarks on the gluonic sector, the temperature argument of $a$ and $b$ is replaced by an effective Yang-Mills temperature $T_{\mathrm{YM}}$. The glue temperature of the full theory $T_{\mathrm{glue}}$ is identified with the matter temperature $T$, and $T_{\mathrm{YM}}$ follows from the mapping \cite{Haas:2013}
	\begin{equation}
		\frac{T_{\mathrm{YM}} - T_0^{\mathrm{YM}}}{T_0^{\mathrm{YM}}} = 0.57 \left(\frac{T_{\mathrm{glue}} - T_0^{\mathrm{glue}}}{T_0^{\mathrm{glue}}}\right),
		\label{eq:TYM_Tglue}
	\end{equation}
	with $T_0^{\mathrm{YM}} = T_0^{\mathrm{glue}} = T_0 = 200$ MeV.
	
	\subsection{Fermion vacuum contribution}
	\label{sec:vacuum}
	
	The zero-temperature one-loop quark contribution is
	\begin{equation}
		\Omega_{\bar{q}q}^{\mathrm{vac}}
		= -2 N_c \sum_{i=u,d,s}
		\int_0^{\infty} \frac{d^3 k}{(2\pi)^3}\,
		E_i^*(k),
		\label{eq:Omega_vac_raw}
	\end{equation}
	which is ultraviolet divergent. Dimensional regularization of Eq. (\ref{eq:Omega_vac_raw}) yields \cite{Skokov:2010_vacuum,Schaefer:2012}
	\begin{equation}
		\Omega_{\bar{q}q}^{\mathrm{vac}}
		= -\frac{N_c}{8\pi^2}\sum_{i=u,d,s}
		m_i^{*4}\ln\!\frac{m_i^*}{\Lambda},
		\label{eq:Omega_vac}
	\end{equation}
	where $\Lambda = 600$ MeV is the regularization scale. A change in $\Lambda$ is absorbed by refitting the scalar-sector parameters to the same vacuum observables.
	
	Because $m_i^*$ depends on the scalar fields through Eq. (\ref{eq:mstar}), $\Omega_{\bar{q}q}^{\mathrm{vac}}$ modifies the curvature of the effective potential in field space and contributes to the scalar gap equations derived in Sec. \ref{sec:gap_eqs}. For comparison, we consider a no-sea variant (vac=0) in which $\Omega_{\bar{q}q}^{\mathrm{vac}}$ is dropped, and the mesonic parameters are refitted to the same vacuum observables listed in Sec. \ref{sec:lagrangian}. The two parameter sets are listed in Table \ref{tab:params}.

	\subsection{Gap equations}
	\label{sec:gap_eqs}
	
	The equilibrium mean fields at given $T$, $\mub$, $\muq$, $\mus$ are obtained by minimizing $\Omega$ with respect to all dynamical variables,
	\begin{equation}
		\frac{\partial\Omega}{\partial\varphi_a} = 0,
		\qquad
		\varphi_a \in \{\sigma, \zeta, \delta, \chi, \omega, \rho, \phi, \Ploop, \Ploopbar\}.
		\label{eq:gap}
	\end{equation}
	The gap equations read
	\begin{align}
		\frac{\partial\Omega}{\partial\sigma}
		&= k_0\chi^2\sigma
		- 4 k_1(\sigma^2 + \zeta^2 + \delta^2)\sigma
		- 2 k_2(\sigma^3 + 3\sigma\delta^2)
		\nonumber \\
		&\quad - 2 k_3\chi\sigma\zeta
		- \frac{d}{3}\chi^4\frac{2\sigma}{\sigma^2 - \delta^2}
		+ \frac{\chi^2}{\chi_0^2}\,h_x
		\nonumber \\
		&\quad - \frac{\chi^2}{\chi_0^2} m_\omega\omega^2 \frac{\partial m_\omega}{\partial\sigma}
		- \frac{\chi^2}{\chi_0^2} m_\rho\rho^2 \frac{\partial m_\rho}{\partial\sigma}
		\nonumber \\
		&\quad - \sum_{i=u,d} g_\sigma^i\,\rho_i^s
		= 0,
		\label{eq:gap_sigma}
	\end{align}
	\begin{align}
		\frac{\partial\Omega}{\partial\zeta}
		&= k_0\chi^2\zeta
		- 4 k_1(\sigma^2 + \zeta^2 + \delta^2)\zeta
		- 4 k_2\zeta^3
		\nonumber \\
		&\quad - k_3\chi(\sigma^2 - \delta^2)
		- \frac{d}{3}\frac{\chi^4}{\zeta}
		+ \frac{\chi^2}{\chi_0^2}\,h_y
		\nonumber \\
		&\quad - \frac{\chi^2}{\chi_0^2} m_\phi\phi^2 \frac{\partial m_\phi}{\partial\zeta}
		- g_\zeta^s\,\rho_s^s
		= 0,
		\label{eq:gap_zeta} \\[4pt]
		\frac{\partial\Omega}{\partial\delta}
		&= k_0\chi^2\delta
		- 4 k_1(\sigma^2 + \zeta^2 + \delta^2)\delta
		- 2 k_2(\delta^3 + 3\sigma^2\delta)
		\nonumber \\
		&\quad + 2 k_3\chi\delta\zeta
		+ \frac{2d}{3}\chi^4\frac{\delta}{\sigma^2 - \delta^2}
		- \sum_{i=u,d} g_\delta^i\,\rho_i^s
		= 0,
		\label{eq:gap_delta} \\[4pt]
		\frac{\partial\Omega}{\partial\chi}
		&= k_0\chi(\sigma^2 + \zeta^2 + \delta^2)
		- k_3(\sigma^2 - \delta^2)\zeta
		\nonumber \\
		&\quad + \chi^3\!\left[1 + \ln\frac{\chi^4}{\chi_0^4}\right]
		+ (4 k_4 - d)\chi^3
		\nonumber \\
		&\quad - \frac{4 d}{3}\chi^3
		\ln\!\left[\frac{(\sigma^2 - \delta^2)\zeta}{\sigma_0^2\zeta_0}
		\cdot\frac{\chi^3}{\chi_0^3}\right] + \frac{2\chi}{\chi_0^2}(h_x\sigma + h_y\zeta)
		\nonumber \\
		&\quad - \frac{\chi}{\chi_0^2}(m_\omega^2\omega^2 + m_\rho^2\rho^2 + m_\phi^2\phi^2)
		= 0,
		\label{eq:gap_chi} \\[4pt]
		\frac{\partial\Omega}{\partial\omega}
		&= \frac{\chi^2}{\chi_0^2}\,m_\omega^2\,\omega
		+ 4 g_4\omega^3 + 12 g_4\omega\rho^2
		- \sum_{i=u,d} g_\omega^i\,\rho_i
		= 0,
		\label{eq:gap_omega} \\[4pt]
		\frac{\partial\Omega}{\partial\rho}
		&= \frac{\chi^2}{\chi_0^2}\,m_\rho^2\,\rho
		+ 4 g_4\rho^3 + 12 g_4\omega^2\rho
		- \sum_{i=u,d} g_\rho^i\,\rho_i
		= 0,
		\label{eq:gap_rho} \\[4pt]
		\frac{\partial\Omega}{\partial\phi}
		&= \frac{\chi^2}{\chi_0^2}\,m_\phi^2\,\phi
		+ 8 g_4\phi^3
		- g_\phi^s\,\rho_s
		= 0,
		\label{eq:gap_phi}\\[4pt]
		\frac{\partial\Omega}{\partial\Ploop}
		&= -T^4\!\left[\frac{a(T)}{2}\Ploopbar
		+ \frac{6 b(T)\bigl(\Ploopbar - 2\Ploop^2 + \Ploopbar^2\Ploop\bigr)}{D}\right]
		\nonumber \\
		&\quad - \sum_{i=u,d,s}\frac{2 T N_c}{(2\pi)^3}\int_0^{\infty}\!\! d^3 k
		\left[\frac{e^{-E_-}}{\mathcal{N}_i^-} + \frac{e^{-2 E_+}}{\mathcal{N}_i^+}\right]
		= 0,
		\label{eq:gap_Phi} \\[4pt]
		\frac{\partial\Omega}{\partial\Ploopbar}
		&= -T^4\!\left[\frac{a(T)}{2}\Ploop
		+ \frac{6 b(T)\bigl(\Ploop - 2\Ploopbar^2 + \Ploop^2\Ploopbar\bigr)}{D}\right]
		\nonumber \\
		&\quad - \sum_{i=u,d,s}\frac{2 T N_c}{(2\pi)^3}\int_0^{\infty}\!\! d^3 k
		\left[\frac{e^{-2 E_-}}{\mathcal{N}_i^-} + \frac{e^{-E_+}}{\mathcal{N}_i^+}\right]
		= 0,
		\label{eq:gap_Phibar}
	\end{align}
	with
	\begin{equation}
		D \equiv 1 - 6\Ploop\Ploopbar + 4(\Ploop^3 + \Ploopbar^3) - 3(\Ploop\Ploopbar)^2
		\label{eq:Dhaar}
	\end{equation}
	the Haar-measure argument appearing in Eq. (\ref{eq:U_poly}).
	The scalar density $\rho_i^s$ of flavor $i$ is
	\begin{align}
		\rho_i^s &= -\frac{N_c}{2\pi^2}\,m_i^{*3}\!\left(\ln\frac{m_i^*}{\Lambda} + \frac{1}{4}\right) \nonumber \\
		&+ \gamma_i N_c \int_0^{\infty}\frac{d^3 k}{(2\pi)^3}\,
		\frac{m_i^*}{E_i^*(k)}\bigl(f_i + \bar f_i\bigr),
		\label{eq:rho_s}
	\end{align}
	and the net quark number density of flavor $i$
	\begin{equation}
		\rho_i = \gamma_i N_c \int_0^{\infty}\frac{d^3 k}{(2\pi)^3}\,
		\bigl(f_i - \bar f_i\bigr),
		\label{eq:rho_v}
	\end{equation}
	with Polyakov-loop-modified distributions
	\begin{align}
		f_i &= \frac{\Ploop\,e^{-E_-} + 2\Ploopbar\,e^{-2 E_-} + e^{-3 E_-}}{\mathcal{N}_i^-},
		\label{eq:fi} \\[4pt]
		\bar f_i &= \frac{\Ploopbar\,e^{-E_+} + 2\Ploop\,e^{-2 E_+} + e^{-3 E_+}}{\mathcal{N}_i^+}.
		\label{eq:fbar}
	\end{align}
	The first term in Eq. (\ref{eq:rho_s}) is the vacuum contribution and vanishes in the vac=0 variant. At $g_v = 0$, the source terms in Eqs. (\ref{eq:gap_omega})-(\ref{eq:gap_phi}) vanish and the remaining homogeneous equations admit only the trivial solution $\omega = \rho = \phi = 0$. The effective chemical potentials in Eq. (\ref{eq:mustar}) then reduce to the bare quark chemical potentials, $\mu_i^* = \mu_i$. The effect of nonzero $g_v$ on the finite-$\mu_B$ results is examined in Sec. \ref{sec:gv_sensitivity}.
	
	\subsection{Conserved-charge susceptibilities}
	\label{sec:chi_def}
	The generalized susceptibilities of conserved charges are defined as derivatives of the scaled pressure with respect to the reduced chemical potentials $\hat{\mu}_X \equiv \mu_X/T$ \cite{Bazavov:2012,Borsanyi:2012},
	\begin{equation}
		\chi_{ijk}^{BQS} = \frac{\partial^{i+j+k}(p/T^4)}{\partial\hat{\mu}_B^{\,i}\,\partial\hat{\mu}_Q^{\,j}\,\partial\hat{\mu}_S^{\,k}},
		\label{eq:chi_def}
	\end{equation}
	where $p = -\Omega$. Diagonal susceptibilities are denoted $\chiB{n} \equiv \chi_{n00}^{BQS}$, $\chiQ{n} \equiv \chi_{0n0}^{BQS}$, and $\chiS{n} \equiv \chi_{00n}^{BQS}$. Off-diagonal correlators between two charge sectors are denoted $\chioff{ij}{XY}$, with $X, Y \in \{B, Q, S\}$ and $i+j$ equal to the total order. Mixed three-sector correlators retain the full $\chi_{ijk}^{BQS}$ notation. At the stationary point of $\Omega$, first derivatives with respect to $\mu_X$ reduce to partial derivatives at fixed fields \cite{Chatterjee:2012}. At second and higher orders, the implicit response of the mean fields to chemical-potential variation contributes and must be tracked through Eqs. (\ref{eq:gap_sigma})-(\ref{eq:gap_Phibar}).
	
	In the high-temperature limit, where all flavors become massless and noninteracting, the second-order susceptibilities approach their Stefan-Boltzmann values
	\begin{equation}
		\chi_2^X\big|_{\SB} = \sum_{i=u,d,s} X_i^2,
		\label{eq:SB_limit}
	\end{equation}
	yielding $\chiB{2}|_{\SB} = 1/3$, $\chiQ{2}|_{\SB} = 2/3$, and $\chiS{2}|_{\SB} = 1$. Stefan-Boltzmann limits for the higher-order and off-diagonal quantities computed in this work are collected in Appendix \ref{app:SB}.
	
	\section{Results at \texorpdfstring{$\mu_B = 0$}{muB = 0}}
	\label{sec:results_mub0}
	
	Throughout this section, results are reported for the vacuum-included (vac=1) and the independently refitted no-sea (vac=0) parameter sets of Table \ref{tab:params}. Solid curves denote vac=1 and dashed curves vac=0 in every figure. The diagonal susceptibilities through eighth order and the complete fourth-order off-diagonal tensor are evaluated with derivatives in $\mub$, $\muq$, and $\mus$ taken at $\mub = \muq = \mus = 0$.

	\subsection{Order parameters and bulk thermodynamics}
	\label{sec:order_params}
	
	\begin{figure*}[t]
		\centering
		\includegraphics[width=0.80\textwidth]{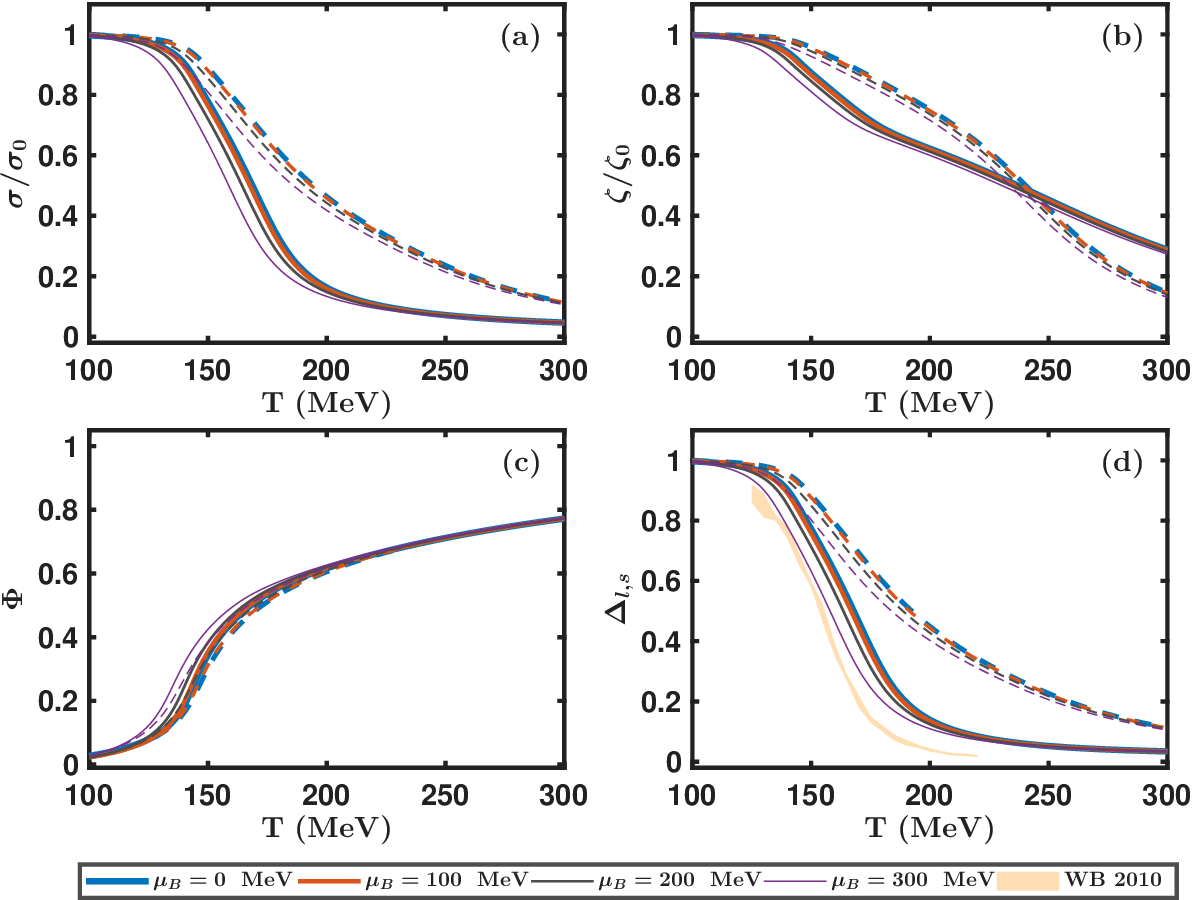}
		\caption{Order parameters vs temperature at $\mub = 0$, 100, 200, and 300 MeV. (a) Non-strange scalar field $\sigma/\sigma_0$, (b) strange scalar field $\zeta/\zeta_0$, (c) Polyakov loop $\Ploop$, (d) subtracted condensate $\Dls$. Each $\mub$ slice has its own color, and thicker lines mark lower $\mub$. Solid (dashed) lines correspond to PCQMF with (without) the fermion vacuum term. The shaded band in panel (d) is the WB 2010 \cite{Borsanyi:2010WB} continuum-extrapolated lattice band for $\Dls$.}
		\label{fig:order_params}
	\end{figure*}
	
	Figure \ref{fig:order_params} shows the chiral and deconfinement order parameters as functions of temperature at $\mu_B = 0$, 100, 200, and 300 MeV. The non-strange scalar field $\sigma/\sigma_0$ in panel (a) undergoes a smooth crossover that shifts to lower temperature with increasing $\mu_B$, and the vac=1 drop is steeper than the vac=0 drop at every $\mu_B$ slice. The strange scalar field $\zeta/\zeta_0$ in panel (b) shifts similarly to lower temperature with increasing $\mu_B$, but the four slices stay much closer together than the $\sigma/\sigma_0$ slices because $\mu_s = \mu_B/3$ remains small compared to $m_s^*$. The vac=1 drop is again steeper than vac=0 at every slice. At fixed $\mu_B$, $\zeta/\zeta_0$ falls more slowly with $T$ than $\sigma/\sigma_0$ because $m_s^*$ is heavier than $m_u^* = m_d^*$. The Polyakov loop $\Phi$ in panel (c) rises through the crossover region with weak $\mu_B$ dependence in both variants. Because the Polyakov-loop potential $\mathcal{U}(\Phi, \bar{\Phi}, T)$ carries no explicit chemical-potential dependence, $\mu_B$ enters $\Phi$ only indirectly through the scalar fields in the gap equations. Panel (d) shows the subtracted chiral condensate $\Delta_{l,s}$ as a function of temperature. In the chiral limit, the quark condensate vanishes at the chiral transition. At a nonzero quark mass, it does not, and a subtraction is used to remove the explicit mass contribution. The lattice definition reads \cite{Bazavov:2009}
	\begin{equation}
		\Delta_{l,s}(T) = \frac{\langle\bar\psi\psi\rangle_{l,T} - (\hat m_l/\hat m_s)\,\langle\bar\psi\psi\rangle_{s,T}}{\langle\bar\psi\psi\rangle_{l,0} - (\hat m_l/\hat m_s)\,\langle\bar\psi\psi\rangle_{s,0}}.
		\label{eq:Dls_lat}
	\end{equation}
	In the PCQMF model, $\sigma$ and $\zeta$ replace the light and strange condensates and $r = h_x/h_y$ replaces the bare-mass ratio \cite{Schaefer:2010_thermo, Singh:2025}, giving
	\begin{equation}
		\Delta_{l,s}(T) = \frac{\sigma - r\zeta}{\sigma_0 - r\zeta_0}.
		\label{eq:Dls}
	\end{equation}
	$\Delta_{l,s}(0) = 1$ by construction, and $\Delta_{l,s}$ decreases with temperature. Both variants lie above the WB 2010 continuum-extrapolated band \cite{Borsanyi:2010WB} across the crossover region. The vac=1 curve sits closer to the band.
	
	\begin{figure*}[t]
		\centering
		\includegraphics[width=0.80\textwidth]{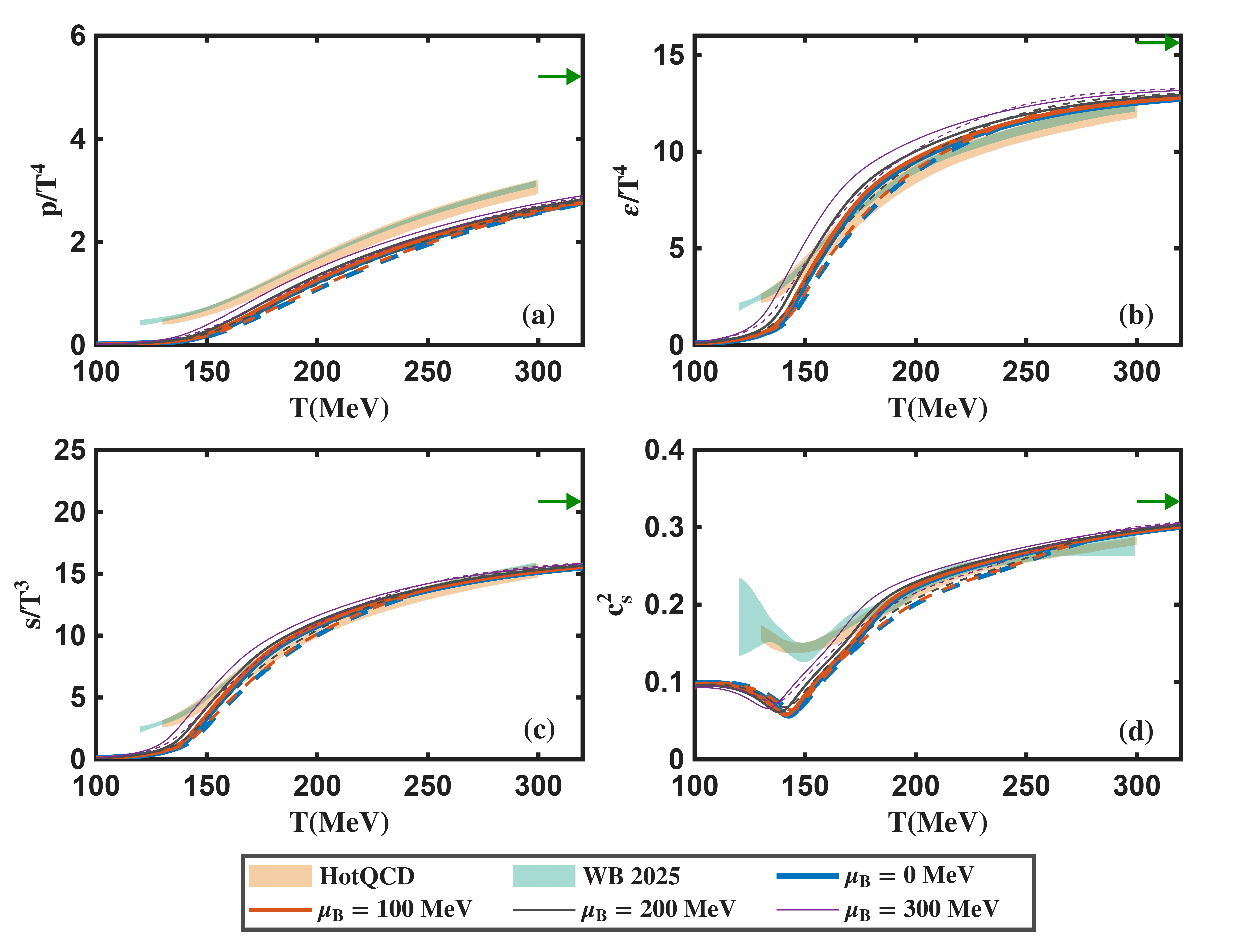}
		\caption{Bulk thermodynamics at $\mub = 0$, 100, 200, and 300 MeV. (a) Pressure $p/T^4$, (b) energy density $\varepsilon/T^4$, (c) entropy density $s/T^3$, (d) speed of sound squared $c_s^2$. Each $\mub$ slice has its own color, and thicker lines mark lower $\mub$. Solid (dashed) lines correspond to PCQMF with (without) the fermion vacuum term. Lattice bands are HotQCD 2014 \cite{HotQCD:2014_EOS} (orange) and Borsanyi et al.\ 2025 \cite{Borsanyi:2025_continuum} (teal). Forest-green right-edge arrows mark the SB limits (Appendix \ref{app:SB}).}
		\label{fig:thermo}
	\end{figure*}
	
	In Fig. \ref{fig:thermo}, the scaled pressure $p/T^4$, energy density $\varepsilon/T^4$, entropy density $s/T^3$, and squared speed of sound $c_s^2$ are compared with continuum-extrapolated lattice data \cite{HotQCD:2014_EOS, Borsanyi:2025_continuum}. The PCQMF model treats quarks as the active degrees of freedom at all temperatures and underpredicts these quantities at low temperatures, where the lattice EoS is dominated by hadronic contributions. At higher temperatures, the model values of $\varepsilon/T^4$, $s/T^3$, and $c_s^2$ lie within $\sim 5\%$ of the lattice bands, while $p/T^4$ remains $15$-$25\%$ below. All four quantities approach the Stefan-Boltzmann (SB) limit from below at high temperature. The minimum value of the speed of sound is $0.057$ (vac=1) and $0.062$ (vac=0), well below the lattice value of $0.14$.
	
	\begin{figure*}[t]
		\centering
		\includegraphics[width=\textwidth]{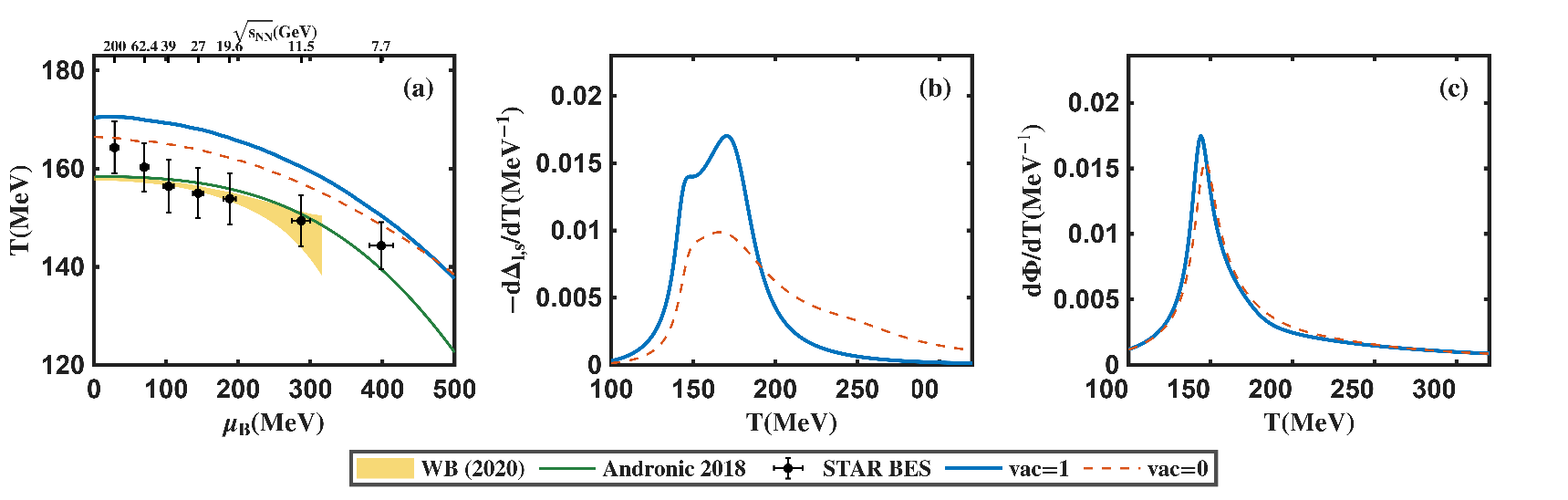}
		\caption{(a) Pseudocritical temperature $\Tpc(\mub)$ from PCQMF gap-equation solutions for $\mub \leq 500$ MeV. (b) Temperature derivative of the subtracted condensate $-d\Dls/dT$ at $\mub = 0$ MeV. (c) Temperature derivative of the Polyakov-loop expectation value $d\Phi/dT$ at $\mub = 0$ MeV. Solid (dashed) lines correspond to PCQMF with (without) the fermion vacuum term. Panel (a) also shows the WB 2020 \cite{Borsanyi:2020WB_PRL} continuum-extrapolated chiral crossover band (gold), the Andronic et al.\ 2018 \cite{Andronic:2018} chemical freeze-out parametrization (green), and STAR BES-I freeze-out points \cite{Adamczyk:2017_freezeout} (black markers with error bars). Upper-axis tick marks give $\sqrt{s_{NN}}$ for the STAR points.}
		\label{fig:phase_diagram}
	\end{figure*}
	
	Figure \ref{fig:phase_diagram}(a) shows the pseudocritical line $\Tpc(\mub)$ for $\mub \leq 500$ MeV. The chiral pseudocritical temperature $\Tpc$ is the temperature at which $-d\Delta_{l,s}/dT$ peaks (Fig. \ref{fig:phase_diagram}(b)), and the deconfinement pseudocritical temperature $\Tdec$ is the temperature at which $d\Ploop/dT$ peaks (Fig. \ref{fig:phase_diagram}(c)). At $\mub = 0$ MeV, $\Tpc = 170.5$ MeV (vac=1) and $166.4$ MeV (vac=0), and $\Tdec = 144.4$ MeV (vac=1) and $146.6$ MeV (vac=0). The chiral-deconfinement splitting is $26.1$ MeV for vac=1 and $19.8$ MeV for vac=0. The same glue-to-Yang-Mills mapping of Eq. (\ref{eq:TYM_Tglue}) \cite{Haas:2013} generates the splitting in both variants. Earlier PCQMF works without this mapping \cite{Kumari:2021,Chahal:2022} placed both crossovers near 165 MeV. The model $\Tpc(\mub)$ lies above the continuum-extrapolated chiral crossover band \cite{Borsanyi:2020WB_PRL}. The offset comes from the mean-field $\Tpc(0)$ overestimate. The chemical freeze-out parametrization \cite{Andronic:2018} and the STAR BES-I freeze-out points \cite{Adamczyk:2017_freezeout} are overlaid. Each sits below $\Tpc(\mub)$ because chemical freeze-out occurs after the partonic phase has cooled through the crossover.
	
	Figure \ref{fig:phase_diagram}(b) shows the temperature derivative of the subtracted condensate, $-d\Dls/dT$, as a function of $T$ at $\mub = 0$ MeV for both variants. The vac=1 curve rises from below, inflects near $\Tdec$, and continues into the main chiral peak at $\Tpc$. The vac=0 curve rises monotonically into its chiral peak with no such inflection. The splitting from the glue mapping is 26.1 MeV in vac=1 and 19.8 MeV in vac=0. The deconfinement-driven inflection in $\Delta_{l,s}(T)$ near $\Tdec$ becomes visible in $-d\Dls/dT$ in vac=1 through the coupled gap equations, and is absent in vac=0. Figure \ref{fig:phase_diagram}(c) shows $d\Phi/dT$ versus $T$ at $\mub = 0$ MeV for both variants. The vac=1 curve peaks at $\Tdec = 144.4$ MeV and the vac=0 curve peaks at $\Tdec = 146.6$ MeV. 
	
	The lattice value $\Tpc = 158.0 \pm 0.6$ MeV \cite{Borsanyi:2020WB_PRL} lies below both variants. The vac=0 result is closer to the lattice, but this is accidental. The vacuum term is required for the correct universality class in the chiral limit \cite{Skokov:2010_vacuum}. The (2+1)-flavor PQM model with the vacuum term \cite{Chatterjee:2012, Chatterjee:2012_thermo} reproduces lattice thermodynamics. PCQMF and PQM with the vacuum term both place $\Tpc$ above the lattice value \cite{Gupta:2012}. The PCQMF $\Tpc$ difference between variants, approximately $4$ MeV, is smaller than the vacuum-term shifts reported for PQM \cite{Skokov:2010_vacuum, Chatterjee:2012_thermo}. PNJL calculations \cite{Bhattacharyya:2010} also place $\Tpc$ above the lattice value. In all three model classes, the overshoot magnitude varies with the parameter set. Adding mesonic fluctuations beyond mean field via the fRG lowers $\Tpc$ toward the lattice value at $\mub = 0$ MeV \cite{Herbst:2011, Fu:2020, Fu:2020_strangeness_phasediag, Fu:2021_hyperorder}. A beyond-mean-field PCQMF treatment of these fluctuations is left for future work.

	\subsection{Second-order susceptibilities}
	\label{sec:chi2}
	
\begin{figure*}
		\centering
		\includegraphics[width=\textwidth]{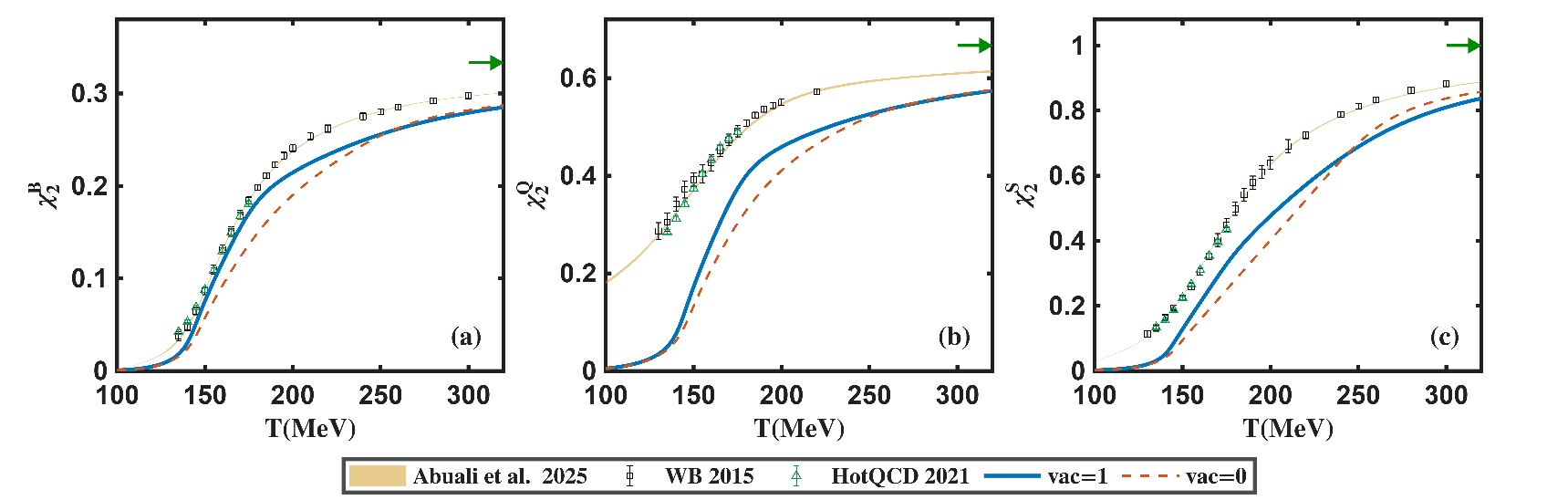}
		\caption{Diagonal second-order susceptibilities (a) Baryon-number $\chiB{2}$, (b) electric-charge $\chiQ{2}$, (c) strangeness $\chiS{2}$ vs temperature at $\mub = 0$ MeV. Solid (dashed) lines correspond to PCQMF with (without) the fermion vacuum term. Lattice data are WB 2015 \cite{Bellwied:2015WB} (black squares), HotQCD 2021 \cite{Bollweg:2021HotQCD} (dark-green triangles), and Abuali et al.\ 2025 \cite{Abuali:2025WB} (amber bands). Forest-green right-edge arrows mark the SB limits (Appendix \ref{app:SB}).}
		\label{fig:chi2}
	\end{figure*}

	Figure \ref{fig:chi2} shows the diagonal second-order susceptibilities $\chiB{2}$, $\chiQ{2}$, and $\chiS{2}$ at $\mub = 0$ MeV. Three lattice datasets are overlaid for comparison \cite{Bellwied:2015WB,Bollweg:2021HotQCD,Abuali:2025WB}. The SB limits $\chiB{2}|_\SB = 1/3$, $\chiQ{2}|_\SB = 2/3$, and $\chiS{2}|_\SB = 1$ are marked by forest-green right-edge arrows. All three model curves lie below the lattice bands across the full temperature range and approach the SB limits from below at high $T$. The deficit arises from the absence of hadronic contributions at low $T$ and the absence of bosonic fluctuations beyond the mean field at high $T$. The same mean-field limitations cause the pressure deficit (Fig. \ref{fig:thermo}). The inflection of $\chiB{2}$ tracks $\Tdec$ rather than $\Tpc$. In vac=1, this inflection lies $\approx 2$ MeV below the vac=0 value, and the rise is $25\%$ steeper.
	
	The strangeness susceptibility $\chiS{2}$ approaches its SB limit more slowly than $\chiB{2}$ and $\chiQ{2}$ because the strange constituent mass remains above the light-quark mass through the crossover region. All three model curves underpredict the lattice at low $T$, but the deficit is most severe for $\chiQ{2}$ and smallest for $\chiB{2}$. Pions, the lightest charged hadrons, dominate $\chiQ{2}$ in the hadronic phase but are frozen at their mean-field classical value in PCQMF and do not contribute to the fluctuations. The same deficit was identified in PQM with the vacuum term \cite{Chatterjee:2012}. The $\chiB{2}$ and $\chiS{2}$ deficits are smaller because their dominant hadronic carriers are nucleons and kaons, both much heavier than pions. The same hierarchy appears in PNJL \cite{Bhattacharyya:2010} and NJL \cite{Fan:2017} mean-field calculations.
	
\begin{figure*}[t]
		\centering
		\includegraphics[width=\textwidth]{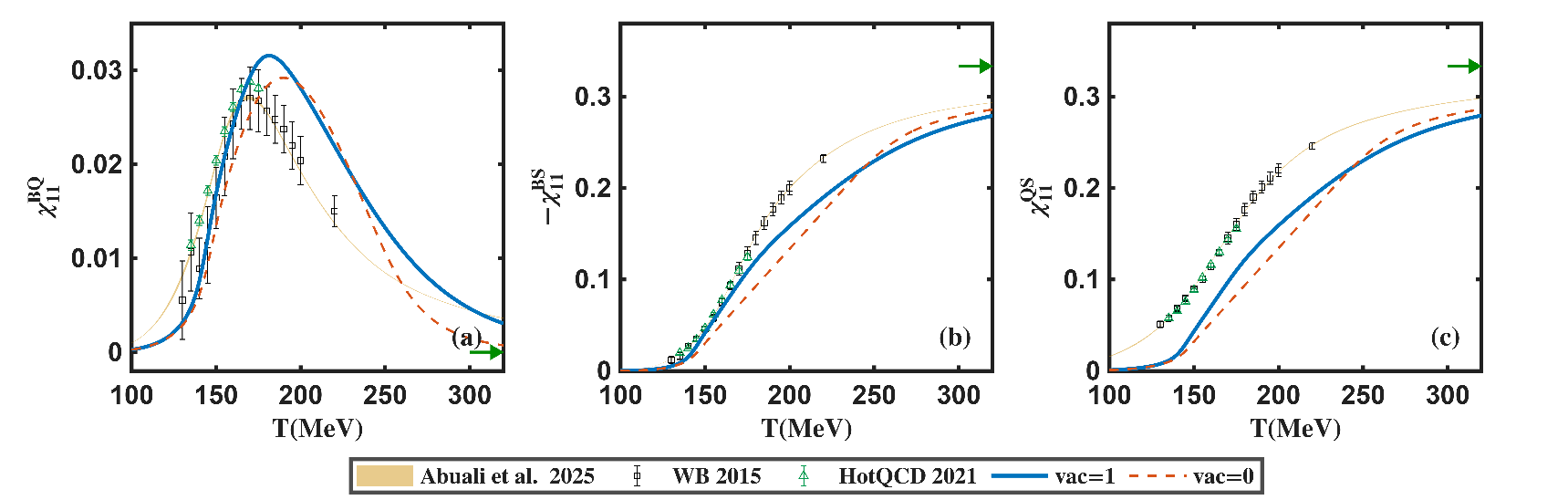}
		\caption{Second-order off-diagonal correlators (a) Baryon-charge $\chioff{11}{BQ}$, (b) baryon-strangeness $-\chioff{11}{BS}$, (c) charge-strangeness $\chioff{11}{QS}$ vs temperature at $\mub = 0$ MeV. Solid (dashed) lines correspond to PCQMF with (without) the fermion vacuum term. Lattice data as in Fig. \ref{fig:chi2}. Forest-green right-edge arrows mark the SB limits (Appendix \ref{app:SB}).}
		\label{fig:offdiag2}
	\end{figure*}

	Figure \ref{fig:offdiag2} shows second-order off-diagonal correlators at $\mub = 0$ MeV. Panel (b) plots $-\chioff{11}{BS}$ because $\chioff{11}{BS}$ is negative across the full $T$ range. In the hadronic phase, hyperons carry $B = +1$ and $S < 0$. In the partonic phase, the strange quark carries $B = +1/3$ and $S = -1$. Hence, $\chioff{11}{BS}$ remains negative in both regimes. $\chioff{11}{BQ}$ peaks near the crossover and decays toward zero on both sides. At low $T$, the relevant $(B, Q)$ degrees of freedom are heavy and suppress the correlator. At high $T$, the up-quark contribution to $\chioff{11}{BQ}$ is canceled by the combined down and strange contributions, so the correlator vanishes in the SB limit \cite{Chatterjee:2012}. By contrast, $-\chioff{11}{BS}$ and $\chioff{11}{QS}$ rise monotonically toward the SB limit $+1/3$. The strange quark is the only flavor carrying $S$, so its liberation through the crossover drives both correlators upward. $\chioff{11}{BQ}$ in panel (a) tracks the lattice data \cite{Bellwied:2015WB,Bollweg:2021HotQCD,Abuali:2025WB} through the peak. In panels (b) and (c), the lattice data lie above both model variants. Hyperons, which carry both $B$ and $S$, are absent from the PCQMF spectrum, since the model contains only quark and meson degrees of freedom. Kaons, which carry $Q$ and $S$, are frozen at their mean-field classical value.
	
	\subsection{Higher-order baryon susceptibilities}
	\label{sec:chi_higher}
	
\begin{figure*}[t]
		\centering
		\includegraphics[width=\textwidth]{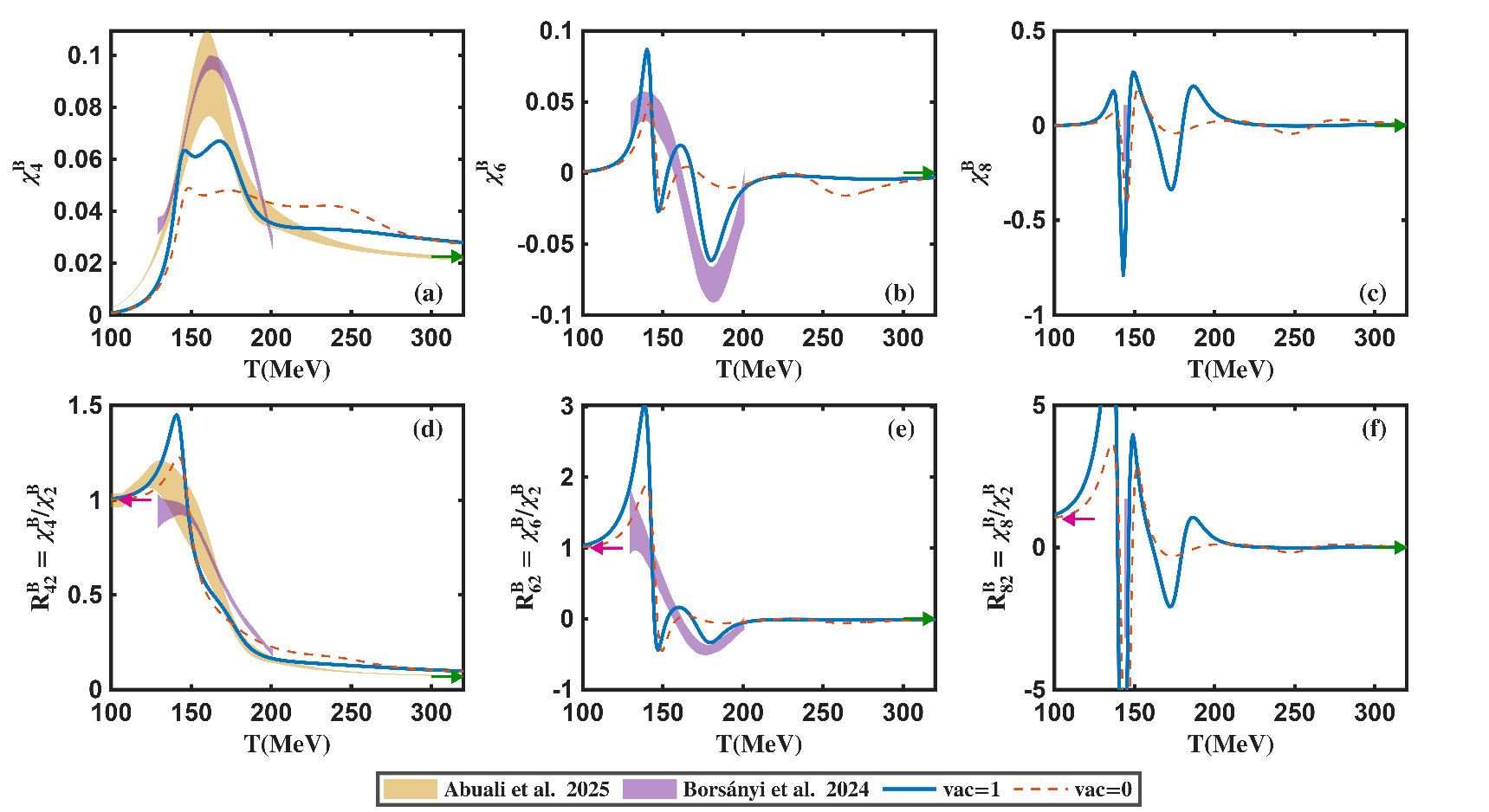}
		\caption{Higher-order diagonal baryon susceptibilities and their ratios at $\mub = 0$ MeV. The top row shows (a) fourth-order $\chiB{4}$, (b) sixth-order $\chiB{6}$, and (c) eighth-order $\chiB{8}$. The bottom row shows the cumulant ratios (d) $R_{42}^B$, (e) $R_{62}^B$, and (f) $R_{82}^B$. Solid (dashed) lines correspond to PCQMF with (without) the fermion vacuum term. Lattice bands are Borsanyi et al.\ 2024 \cite{Borsanyi:2024_chi68} (deep purple) and Abuali et al.\ 2025 \cite{Abuali:2025WB} (amber). Forest-green right-edge arrows mark the SB limits (Appendix \ref{app:SB}). Magenta left-edge arrows in the bottom row mark the HRG baseline $R_{n2}^B = 1$.}
		\label{fig:chi_higher_B}
	\end{figure*}

	Figure \ref{fig:chi_higher_B} shows the higher-order diagonal baryon susceptibilities $\chiB{4}$, $\chiB{6}$, and $\chiB{8}$ (top row) and the cumulant ratios $R_{n2}^B \equiv \chiB{n}/\chiB{2}$ for $n = 4, 6, 8$ (bottom row) at $\mub = 0$ MeV. Continuum lattice bands \cite{Borsanyi:2024_chi68} appear in panels (a), (b), (d), and (e), with additional bands \cite{Abuali:2025WB} in panels (a) and (d). The HRG baseline $R_{n2}^B = 1$ provides a low-temperature reference for the bottom row.
	
	The $\chiB{4}$ curve develops a double-peak structure in both variants. The maxima lie near $\Tdec$ and $\Tpc$. The double peak is the higher-derivative resolution of the chiral-deconfinement splitting. Each crossover contributes its own oscillatory feature to the pressure derivatives, and the splitting separates the two contributions. At higher derivative orders, $\chiB{6}$ and $\chiB{8}$ exhibit multiple zero crossings. Dominant $\chiB{8}$ minima also lie near $\Tdec$ and $\Tpc$. By contrast, the continuum lattice resolves a single crossover region at $\mub = 0$ MeV and shows only a single peak in $\chiB{4}$ and a single zero crossing in $\chiB{6}$ \cite{Borsanyi:2024_chi68}. The additional sign changes in PCQMF result from the resolved chiral-deconfinement splitting absent in the lattice. The dominant negative dip of $\chiB{6}$ across the crossover is common to PCQMF and lattice. This dip is the universal $\chiB{6}$ sign change predicted by $O(4)$ scaling \cite{Friman:2011}. 
	
	In Fig. \ref{fig:chi_higher_B}(d), both PCQMF curves of $R_{42}^B$ start at the HRG baseline at low temperature. They rise to a single peak in the crossover region and decline toward the SB limit $6/(9\pi^2)$ at high temperature. The double peak of $\chiB{4}$ is suppressed in the ratio because $\chiB{2}$ is a smoothly rising denominator that absorbs the shallow dip between the two peaks. The lattice $R_{42}^B$ shows a single peak \cite{Borsanyi:2024_chi68,Abuali:2025WB}. The vac=0 curve sits within the lattice envelope, and the vac=1 curve overshoots it. In Fig. \ref{fig:chi_higher_B}(e) and (f), the PCQMF $R_{62}^B$ and $R_{82}^B$ carry the multi-crossing structure of $\chiB{6}$ and $\chiB{8}$. The lattice $R_{62}^B$ in panel (e) has a single sign change \cite{Borsanyi:2024_chi68}. The PCQMF $R_{62}^B$ peak overshoots the lattice peak. Panel (f) shows the $R_{82}^B$ oscillation across the crossover region. The vac=1 amplitudes substantially exceed vac=0. Large amplitudes for higher-order baryon ratios also appear in the three-flavor PQM model with $R_{n2}^B$ extended to twelfth order \cite{Schaefer:2012}, in the three-flavor PNJL model with quark-number cumulants extended to eighth order \cite{Bhattacharyya:2010}, and in the three-flavor NJL model with $\chiB{n}$ extended to eighth order \cite{Fan:2019}. An fRG calculation that resums quantum, thermal, and density fluctuations recovers the lattice $R_{42}^B$ and $R_{62}^B$ quantitatively at $\mub = 0$ MeV \cite{Fu:2021_hyperorder}.
	
	\subsection{Charge and strangeness higher-order susceptibilities}
	\label{sec:chi_QS_higher}

	\begin{figure*}[t]
		\centering
		\includegraphics[width=\textwidth]{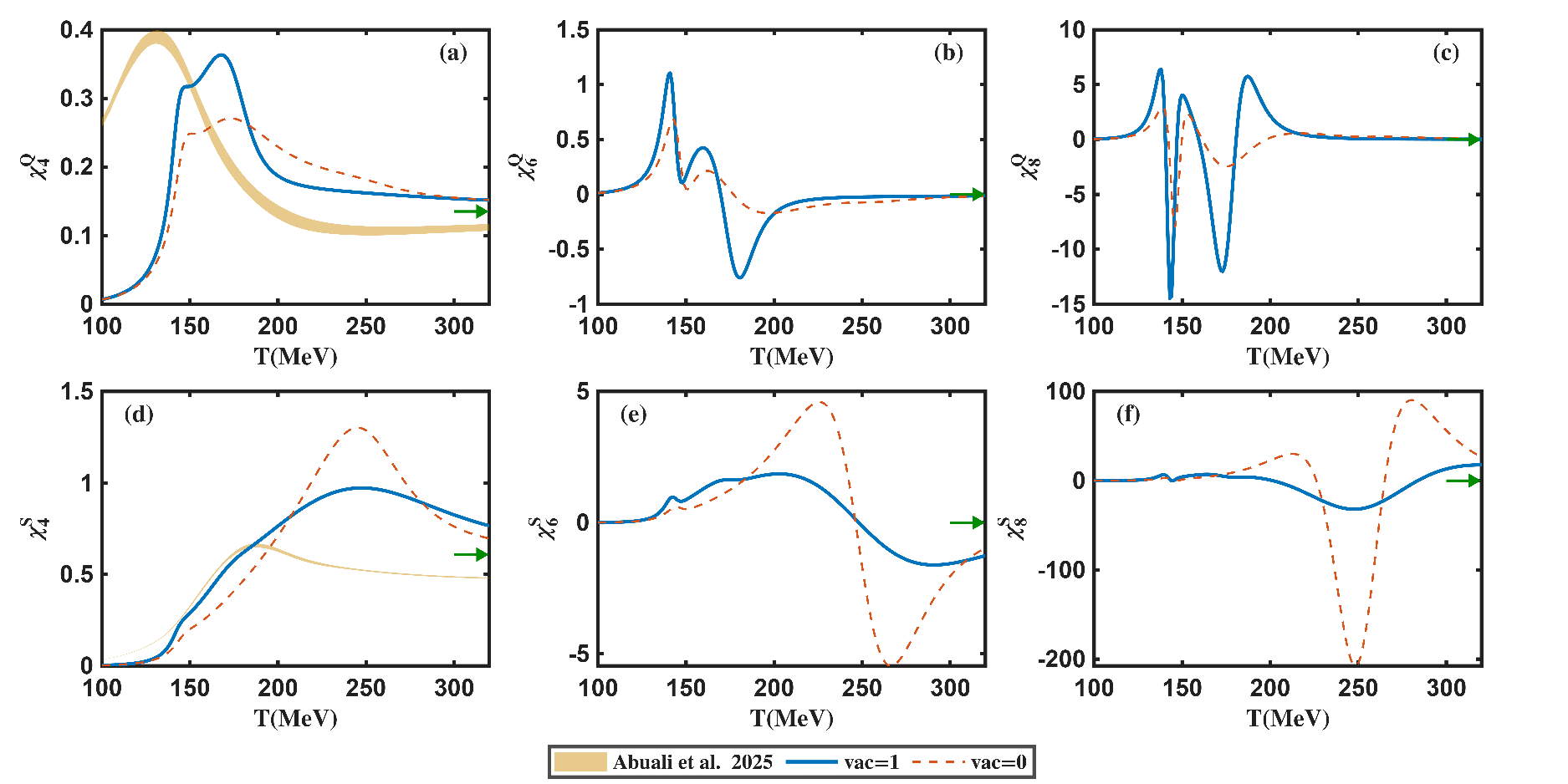}
		\caption{Higher-order diagonal charge and strangeness susceptibilities at $\mub = 0$ MeV. The top row shows (a) fourth-order $\chiQ{4}$, (b) sixth-order $\chiQ{6}$, and (c) eighth-order $\chiQ{8}$. The bottom row shows (d) fourth-order $\chiS{4}$, (e) sixth-order $\chiS{6}$, and (f) eighth-order $\chiS{8}$. Solid (dashed) lines correspond to PCQMF with (without) the fermion vacuum term. Lattice bands in panels (a) and (d) are Abuali et al.\ 2025 \cite{Abuali:2025WB} (amber). Forest-green right-edge arrows mark the SB limits (Appendix \ref{app:SB}).}
		\label{fig:chi_higher_QS}
	\end{figure*}
	
	Figure \ref{fig:chi_higher_QS} presents the diagonal charge and strangeness susceptibilities $\chiQ{n}$ and $\chiS{n}$ for $n = 4, 6, 8$ at $\mub = 0$ MeV. The fourth-order panels carry continuum-extrapolated lattice bands \cite{Abuali:2025WB}. The sixth- and eighth-order panels are model predictions without continuum lattice benchmarks.
	
	$\chiQ{4}$ exceeds $\chiB{4}$ in peak amplitude. The peak amplitudes track the SB-limit inequality $\sum_f Q_f^4 > \sum_f B_f^4$. The up quark dominates the Q-channel sum, and this flavor hierarchy survives in the crossover region. In Fig. \ref{fig:chi_higher_QS}(a), the vac=1 curve is taller, narrower, and centered at slightly lower temperature than vac=0. At higher $T$, the vac=1 curve matches the lattice peak amplitude \cite{Abuali:2025WB}, while vac=0 underestimates it. In Fig. \ref{fig:chi_higher_QS}(d), $\chiS{4}$ peaks at higher $T$ than $\chiQ{4}$ because the strange chiral condensate melts later than the light condensate. The amplitude ordering of vac=1 and vac=0 reverses for $\chiS{4}$. The vac=0 curve peaks above vac=1, opposite to Fig. \ref{fig:chi_higher_QS}(a). The lattice band peaks at a lower temperature and lower amplitude than the PCQMF curves.
	
	In Fig. \ref{fig:chi_higher_QS}(b), $\chiQ{6}$ oscillates across the chiral crossover region. The free massless quark pressure is a polynomial of degree four in $\mu_X/T$, so $\chi_n^X|_\SB = 0$ for all $n \geq 6$. The vac=1 curve shows two positive peaks separated by a shallow dip. A deep negative trough follows at higher $T$. These two maxima resolve the chiral-deconfinement splitting at sixth-derivative order, the same pattern as in $\chiB{4}$ (Sec. \ref{sec:chi_higher}). The vac=1 amplitudes substantially exceed vac=0. In Fig. \ref{fig:chi_higher_QS}(e), $\chiS{6}$ oscillates at higher $T$, where the strange condensate proportional to $\zeta$ is melting. The vac=1 curve carries the $\chiQ{6}$ chiral-deconfinement signature near $\Tdec$ and $\Tpc$, plus a much larger peak and trough in the strange-melting region. The amplitude ordering reverses for $\chiS{6}$. The vac=0 amplitudes substantially exceed vac=1 in the strange-melting region. The basic sign-change pattern of $\chi_6^{Q, S}$ was reported earlier in (2+1)-flavor PQM with the vacuum term \cite{Chatterjee:2012} and in PCQMF without it \cite{Chahal:2022}. Those calculations used parameter sets where the chiral and deconfinement crossovers nearly coincide.
	
	In Fig. \ref{fig:chi_higher_QS}(c), $\chiQ{8}$ shows two deep negative minima near $\Tdec$ and $\Tpc$. The two minima are the eighth-derivative analog of the $\chiQ{6}$ double maximum. The vac=1 amplitudes substantially exceed vac=0, with the disparity larger than at sixth order. In Fig. \ref{fig:chi_higher_QS}(f), $\chiS{8}$ shows a similar structure to $\chiS{6}$. The vac=1 curve carries a large negative dip in the strange-melting region. As in $\chiS{4}$ and $\chiS{6}$, vac=0 dominates the strange-melting region. The dominance is more pronounced at this order. Earlier eighth-order calculations in effective models covered only the baryon channel \cite{Schaefer:2012,Fan:2019,Chahal:2022}.
	
	\subsection{Complete fourth-order off-diagonal tensor}
	\label{sec:offdiag4}
	
	\begin{figure*}[t]
		\centering
		\includegraphics[width=\textwidth]{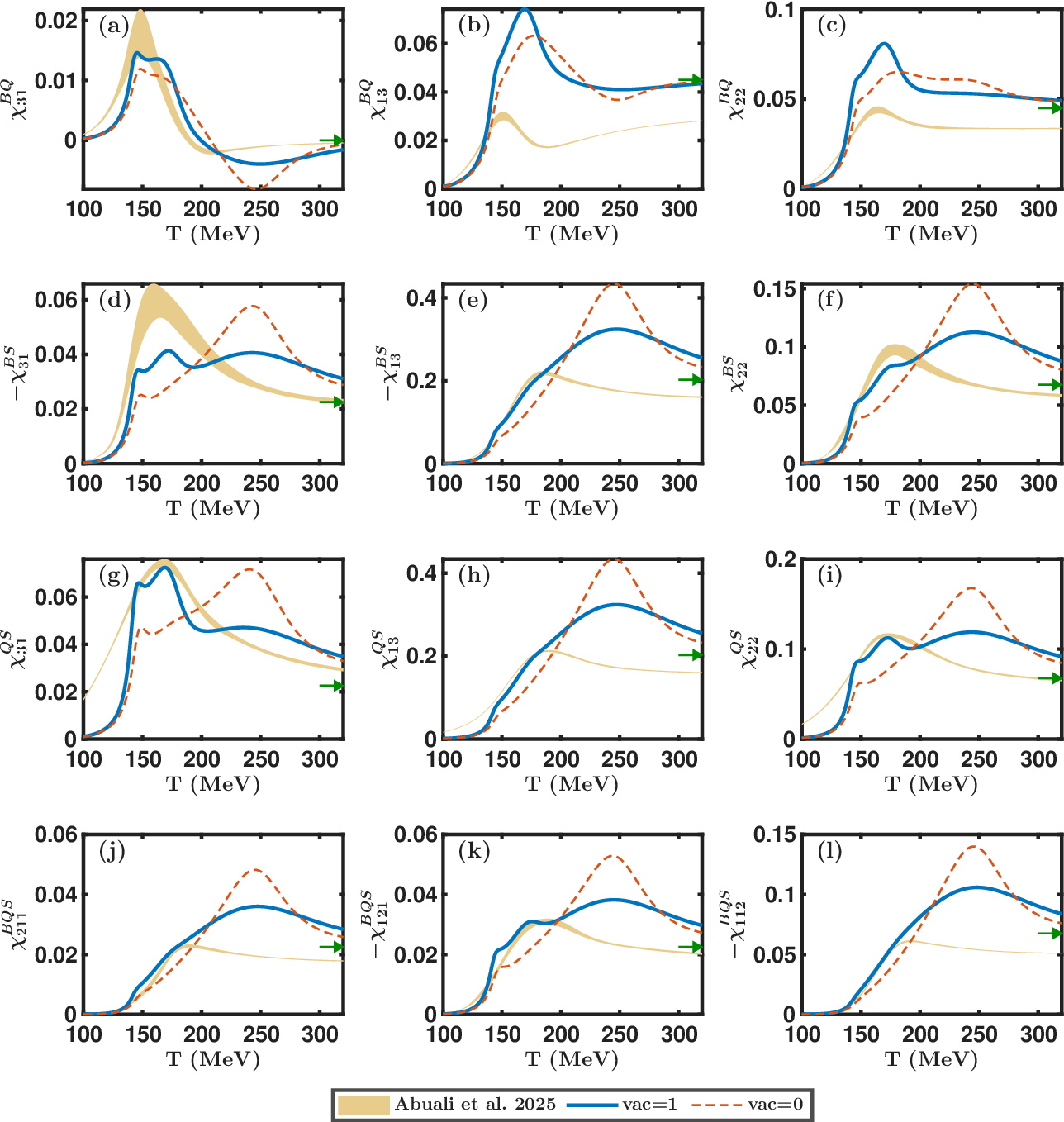}
		\caption{Complete fourth-order off-diagonal BQS tensor vs temperature at $\mub = 0$ MeV. Rows 1-3 show the two-charge correlators $\chioff{ij}{BQ,BS,QS}$ with $i+j=4$, columns ordered as $\chi_{31}$, $\chi_{13}$, $\chi_{22}$. Row 4 shows the mixed correlators $\chioff{ijk}{BQS}$ with $i+j+k=4$, columns ordered as $\chi_{211}$, $-\chi_{121}$, $-\chi_{112}$. Panels (d), (e), (k), (l) carry an overall minus sign to match the convention of Abuali et al.\ 2025 \cite{Abuali:2025WB}. Solid (dashed) lines correspond to PCQMF with (without) the fermion vacuum term. Amber bands are Abuali et al.\ 2025 \cite{Abuali:2025WB}. Forest-green right-edge arrows mark the SB limits (Appendix \ref{app:SB}).}
		\label{fig:offdiag4}
	\end{figure*}
	
	Figure \ref{fig:offdiag4} presents the twelve independent fourth-order off-diagonal correlators at $\mub = 0$ MeV. Rows 1-3 hold the nine two-charge components $\chioff{ij}{BQ,BS,QS}$ with $i+j=4$. Row 4 holds the three mixed components $\chioff{ijk}{BQS}$ with $i+j+k=4$. Panels (d), (e), (k), (l) carry an overall minus sign to match the convention of \cite{Abuali:2025WB}, so that all twelve plotted quantities have positive SB limits. The same twelve correlators have been computed in (2+1)-flavor PQM with the vacuum term at $\mub = 0$ MeV \cite{Chatterjee:2012}. The complete fourth-order off-diagonal tensor in PCQMF with the fermion vacuum term has not been reported elsewhere. The vac=1 peaks exceed vac=0 in the BQ row, and at high $T$ the vac=0 peaks exceed vac=1 in the BS, QS, and BQS rows.
	
	All but one of the twelve plotted quantities remain positive throughout the figure. Panel (a) carries the lone sign change, the SB limit vanishes there, and the curve crosses zero near $T \approx 180$ MeV. The cancellation follows from $\sum_f Q_f = 2/3 - 1/3 - 1/3 = 0$ across the $(u, d, s)$ triplet. The three baryon derivatives contribute only a flavor-independent prefactor $(1/3)^3$. The high-$T$ residual in panel (a) therefore probes quark-mass effects rather than free-quark counting. $\chioff{22}{BQ}$ and $\chioff{13}{BQ}$ carry comparable vac=1 peaks of $0.081$ and $0.074$, well above $\chioff{31}{BQ}$ at $0.015$. Peaks in panels (b) and (c) track the light chiral crossover at $T \approx 170$ MeV. The peak in panel (a) shifts to $T \approx 145$ MeV, near $\Tdec$. The vac=1 peaks exceed vac=0 by about $20\%$ across the BQ row. In panels (b) and (c), the lattice central values sit well below the model peaks. Panel (a) shows the closest lattice-model agreement in the BQ row. The band overlaps the vac=1 peak in temperature and amplitude.
	
	$\chioff{13}{BS}$ is the dominant BS correlator, well above $\chioff{22}{BS}$ and $\chioff{31}{BS}$ in peak amplitude. Dominant peaks of panels (e) and (f) lie at $T \approx 245$ MeV, in the strange-melting region. Panels (d) and (f) carry a double-peak fine structure across $T \approx 150$-$170$ MeV that resolves the chiral-deconfinement splitting. Vac=0 peaks exceed vac=1 across the BS row. The lattice band in panel (d) sits above both model variants in the crossover region. In panel (f), the lattice exceeds both model variants in the crossover region, then falls below them at high $T$.
	
	The QS row repeats the BS hierarchy. $\chioff{13}{QS}$ dominates, well above $\chioff{22}{QS}$ and $\chioff{31}{QS}$. Dominant peaks of panels (h) and (i) lie at $T \approx 245$ MeV, in the strange-melting region. Panels (g) and (i) carry a double-peak fine structure across $T \approx 150$-$170$ MeV that resolves the chiral-deconfinement splitting. At high $T$, vac=0 peaks exceed vac=1 across all three panels. At low $T$, the vac=1 peak in panels (g) and (i) exceeds vac=0 and matches the lattice peak in temperature and magnitude. In panel (h), the model overshoots the lattice throughout. 
	
	$\chioff{112}{BQS}$ in panel (l) dominates the BQS row. $\chioff{121}{BQS}$ in panel (k) and $\chioff{211}{BQS}$ in panel (j) are much smaller and comparable to each other. Dominant peaks across the BQS row lie at $T \approx 245$ MeV, in the strange-melting region. The chiral-deconfinement double-peak across $T \approx 150$-$170$ MeV appears in panel (k). The vac=0 peaks exceed vac=1 across the BQS row. In all three panels, the vac=1 curve shows reasonable agreement with the lattice band for $T < 200$ MeV.
	
	The same qualitative pattern appears in (2+1)-flavor PQM with the vacuum term \cite{Chatterjee:2012}. There, the BQ correlators peak sharply at the crossover, and the BS, QS, and BQS correlators rise steeply and melt slowly toward the SB limit. A double peak appears only as a hint in that calculation, in the $\chi_{22}$ and $\chi_{31}$ components of the BS and QS rows and in $\chioff{121}{BQS}$, where the chiral and deconfinement crossovers nearly coincide. The $26$ MeV splitting in PCQMF resolves the same components into a clear double peak. PCQMF also supplies the explicit vacuum-term comparison and the lattice confrontation absent from that work. In the BS, QS, and BQS rows, the model curves overshoot the lattice peak and approach the SB limit from above, while the lattice band approaches from below. The mean-field treatment of mesonic fluctuations places the model peaks above the SB limit. The fRG resums these fluctuations and softens the peaks in the diagonal sector \cite{Fu:2021_hyperorder}.
	
	\subsection{Correlation ratios}
	\label{sec:corr_ratios}
	
	\begin{figure*}[t]
		\centering
		\includegraphics[width=0.80\textwidth]{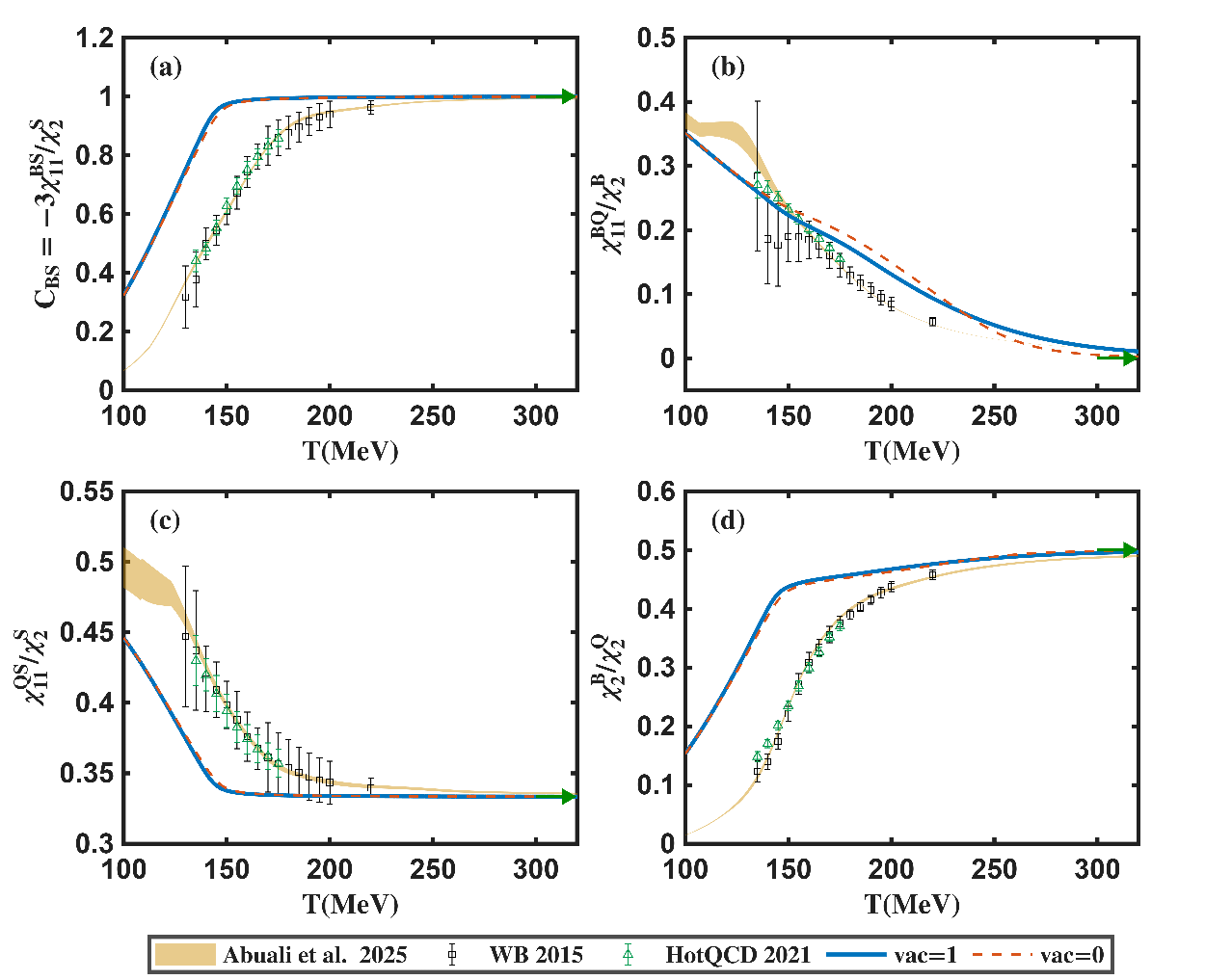}
		\caption{Susceptibility ratios (a) $C_{BS} = -3\chioff{11}{BS}/\chiS{2}$, (b) $\chioff{11}{BQ}/\chiB{2}$, (c) $\chioff{11}{QS}/\chiS{2}$, and (d) $\chiB{2}/\chiQ{2}$ vs temperature at $\mub = 0$ MeV. Solid (dashed) lines correspond to PCQMF with (without) the fermion vacuum term. Lattice data are WB 2015 \cite{Bellwied:2015WB} (black squares), HotQCD 2021 \cite{Bollweg:2021HotQCD} (dark-green triangles), and Abuali et al.\ 2025 \cite{Abuali:2025WB} (amber bands). The Abuali et al.\ bands are constructed by uncorrelated error propagation, which overestimates the true uncertainty. Forest-green right-edge arrows mark the SB limits (Appendix \ref{app:SB}).}
		\label{fig:corr_ratios}
	\end{figure*}
	
	Figure \ref{fig:corr_ratios} shows the temperature dependence of four susceptibility ratios at $\mub = 0$ MeV, namely (a) $C_{BS} = -3\chioff{11}{BS}/\chiS{2}$, (b) $\chioff{11}{BQ}/\chiB{2}$, (c) $\chioff{11}{QS}/\chiS{2}$, and (d) $\chiB{2}/\chiQ{2}$. The ratio $C_{BS}$ serves as a diagnostic of the strangeness carrier across the deconfinement transition \cite{Koch:2005}. Three lattice datasets are overlaid \cite{Bellwied:2015WB,Bollweg:2021HotQCD,Abuali:2025WB}.
	
	In Fig. \ref{fig:corr_ratios}(a), $C_{BS}$ rises monotonically and approaches unity above $T \approx 170$ MeV. The rise tracks the change in strangeness carriers across the crossover. In the hadronic phase, kaons with $B = 0$ dominate $\chiS{2}$ while $\chioff{11}{BS}$ receives contributions only from strange baryons. In the partonic phase, the strange quark contributes to both $\chioff{11}{BS}$ and $\chiS{2}$. The vac=1 and vac=0 curves nearly coincide. Both $\chioff{11}{BS}$ and $\chiS{2}$ are strange-sector quantities, and the ratio divides out their common vacuum-term dependence. Their maximum separation is $\sim 0.01$ near $T = 150$ MeV. The PCQMF rise is steeper than the lattice. At $T = 150$ MeV, the model curves are near unity while the lattice datasets remain near $0.6$.
	
	The remaining three ratios approach their SB limits monotonically. In Fig. \ref{fig:corr_ratios}(b), $\chioff{11}{BQ}/\chiB{2}$ decays from above through the crossover toward zero. The PCQMF curves track the lattice data within errors across the crossover region, unlike panel (a). Within the same range, the vac=0 curve lies above the vac=1 curve. The ordering reverses above $T \approx 250$ MeV. In Fig. \ref{fig:corr_ratios}(c) and (d), the ratios $\chioff{11}{QS}/\chiS{2}$ and $\chiB{2}/\chiQ{2}$ approach their SB limits from above and below, respectively. The vac=1 and vac=0 curves nearly coincide in both panels. The PCQMF approach to SB is sharper than the lattice in both.	The same qualitative ratio behavior is found in (2+1)-flavor PQM with the vacuum term \cite{Chatterjee:2012}. These ratios are accessible through net-particle cumulants measured in heavy-ion collisions, and a comparison along the freeze-out line is left for future work.

	\section{Results at finite \texorpdfstring{$\mu_B$}{muB}}
	\label{sec:results_finite_mub}
	
	In this section, the susceptibilities of Sec. \ref{sec:results_mub0} are extended to finite baryon chemical potential up to $\mub = 500$ MeV. The chemical potentials $\muq$ and $\mus$ are held at zero, which isolates the response of the diagonal susceptibilities to baryon density alone. The vac=1 and vac=0 comparison and the solid/dashed convention of Sec. \ref{sec:results_mub0} carry over to every figure.
	
	\subsection{Second- and fourth-order diagonal susceptibilities at finite $\mub$}
	\label{sec:chi24_finite}
	
	\begin{figure*}[t]
		\centering
		\includegraphics[width=\textwidth]{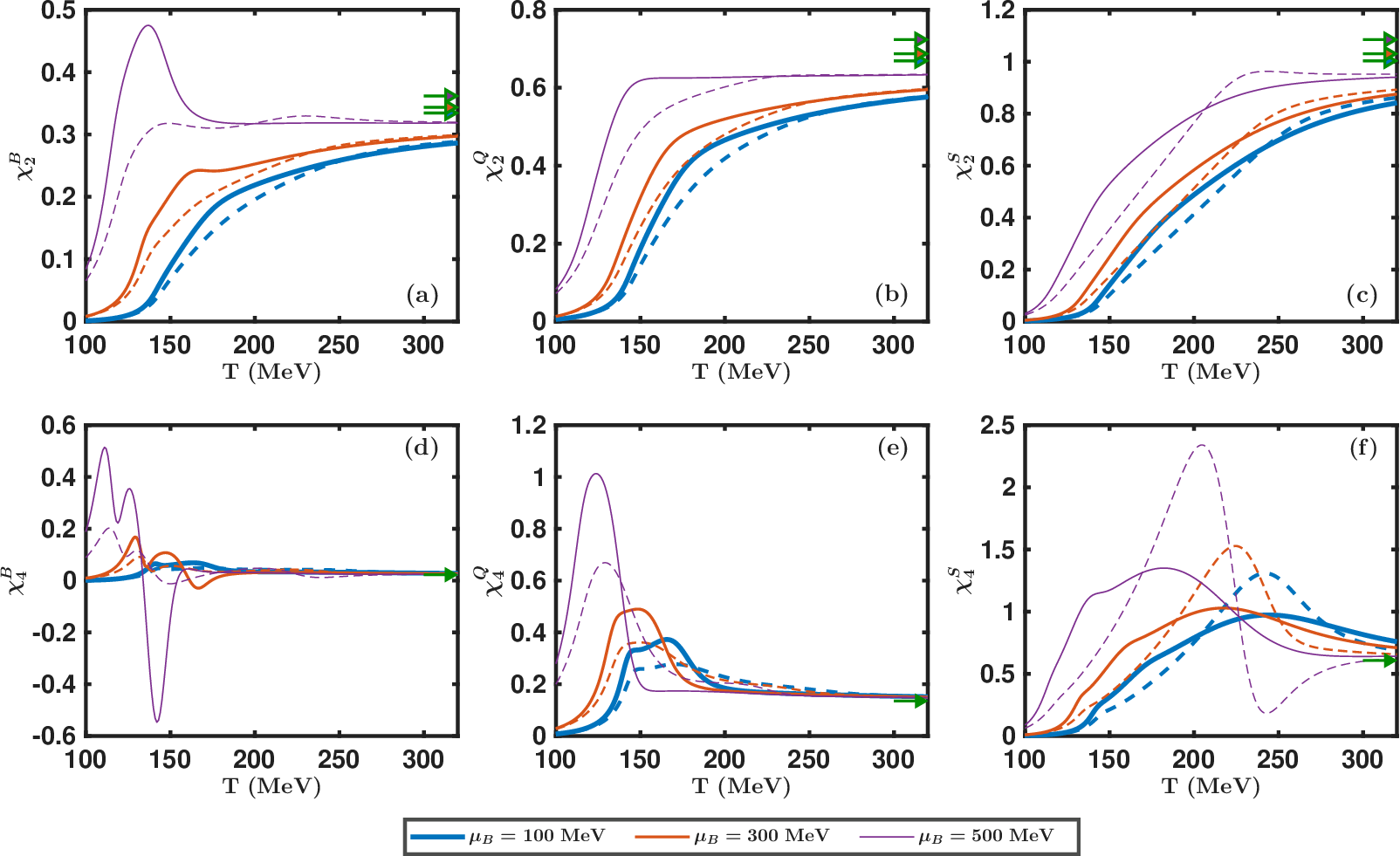}
		\caption{Second- and fourth-order diagonal susceptibilities at finite baryon chemical potential. The top row shows (a) $\chiB{2}$, (b) $\chiQ{2}$, (c) $\chiS{2}$. The bottom row shows (d) $\chiB{4}$, (e) $\chiQ{4}$, (f) $\chiS{4}$. Three slices are overlaid in each panel, $\mub = 100$, 300, and 500 MeV. Each $\mub$ slice has its own color, and thicker lines mark lower $\mub$. Solid (dashed) lines correspond to PCQMF with (without) the fermion vacuum term. Forest-green right-edge arrows mark the SB limits. In the top row, the SB limit is $\mub$-dependent at finite $\mub$, so each slice carries its own arrow with the arrowhead filled by slice color (Appendix \ref{app:SB}).}
		\label{fig:chi24_finitemuB}
	\end{figure*}
	
	Figure \ref{fig:chi24_finitemuB} presents the diagonal susceptibilities at $\mub = 100, 300, 500$ MeV. The top row (panels (a)-(c)) shows $\chiB{2}$, $\chiQ{2}$, $\chiS{2}$ and the bottom row (panels (d)-(f)) shows $\chiB{4}$, $\chiQ{4}$, $\chiS{4}$. Each $\mub$ slice carries a distinct color, and line thickness decreases as $\mub$ grows. SB limits appear as forest-green arrows on the right edge of each panel. 
	
	In Fig. \ref{fig:chi24_finitemuB}(a), the $\chiB{2}$ curves are monotonic at $\mub = 100$ and $300$ MeV. At $\mub = 500$ MeV, the vac=1 curve develops a sharp interior peak at $T \approx 137$ MeV. The peak amplitude reaches $0.475$, well above the SB limit $\chiB{2}\big|_\SB = \frac{1}{3} + \frac{\muhat^2}{9\pi^2}$ (Appendix \ref{app:SB}). The vac=0 curve at the same slice peaks at a smaller amplitude and broader width. In Fig. \ref{fig:chi24_finitemuB}(b), $\chiQ{2}$ stays monotonic across all slices and rises toward the SB limit $\chiQ{2}\big|_\SB = \frac{2}{3} + \frac{2\muhat^2}{9\pi^2}$. No analogous interior peak develops in the Q channel because $\muq = 0$ removes the direct chemical-potential drive. In Fig. \ref{fig:chi24_finitemuB}(c), $\chiS{2}$ is likewise monotonic and stays below $\chiS{2}\big|_\SB = 1 + \frac{\muhat^2}{3\pi^2}$. The strangeness sector responds weakly to $\mub$ at $\mus = 0$, and the relevant crossover scale is set by the strange-quark constituent mass.
	
	In Fig. \ref{fig:chi24_finitemuB}(d), $\chiB{4}$ develops multi-peak structure that strengthens and broadens with $\mub$. At $\mub = 100$ MeV both curves show close double peaks near $T = 141$ and $163$ MeV separated by a shallow trough. At $\mub = 300$ MeV, the vac=1 curve dips below zero between two zero crossings at $T = 162$ and $174$ MeV, with a minimum at $-0.029$. At $\mub = 500$ MeV, the vac=1 peak reaches $0.515$ at $T = 111$ MeV, and the trough drops to $-0.547$ at $T = 142$ MeV. The vac=0 curve follows the same sequence, but the trough reaches only $-0.012$. The double-peak fine structure resolves the chiral-deconfinement splitting established in Sec. \ref{sec:chi_higher}. In Fig. \ref{fig:chi24_finitemuB}(e), $\chiQ{4}$ shows a single dominant peak in each slice that shifts to lower $T$ and grows in amplitude with $\mub$. The inflection near $T \approx 148$ MeV at $\mub = 100$ MeV carries the same chiral-deconfinement signature as the $\chiB{4}$ fine structure. In Fig. \ref{fig:chi24_finitemuB}(f), $\chiS{4}$ shows a single peak that shifts from $T \approx 244$ MeV at $\mub = 100$ MeV to $T \approx 182$ MeV at $\mub = 500$ MeV. The vac=0 peak exceeds the vac=1 peak in amplitude across all slices, and the gap widens with $\mub$. This ordering is the opposite of that in the baryon and charge channels.
	
	Susceptibility amplitudes grow with $\mub$ in every channel, but the fourth-order curves grow far more steeply than the second-order curves. Between $\mub = 100$ and $500$ MeV, $\chiB{2}$ peak amplitudes grow by a factor of $1.7$, while $\chiB{4}$ peak amplitudes grow by a factor of $7.5$. As $\mub$ grows, the peak locations shift to lower $T$ along $\Tpc(\mub)$. The $\mub$ sensitivity decreases from baryon to charge to strangeness at both second and fourth order. The baryon channel receives direct chemical-potential drive, while charge and strangeness respond only through the implicit $\mub$ dependence of the scalar fields and Polyakov loop. The amplitude separation between vac=1 and vac=0 observed at $\mub = 0$ persists into the finite-density region. The vac=1 curves retain their enhanced peak amplitudes for $\chiB{2}$, $\chiB{4}$, and $\chiQ{4}$, while $\chiS{4}$ shows the vac=0 dominance established in Sec. \ref{sec:chi_QS_higher}. The $\chiB{4}$ trough also falls deeper in vac=1. Similar finite-$\mub$ growth of baryon-channel susceptibilities has been reported in PQM \cite{Schaefer:2012}, PNJL \cite{Shao:2018, Ferreira:2018}, and fRG \cite{Fu:2021_hyperorder}. The diagonal $\chi_2$ and $\chi_4$ in all three BQS channels at finite $\mub$ have been computed in NJL at the same constraint surface $\muq = \mus = 0$ \cite{Fan:2017}, which provides the closest content-level comparison to Fig. \ref{fig:chi24_finitemuB}.

	\subsection{Second-order off-diagonal correlators at finite $\mub$}
	\label{sec:chi11_finite}
	
	\begin{figure*}[t]
		\centering
		\includegraphics[width=\textwidth]{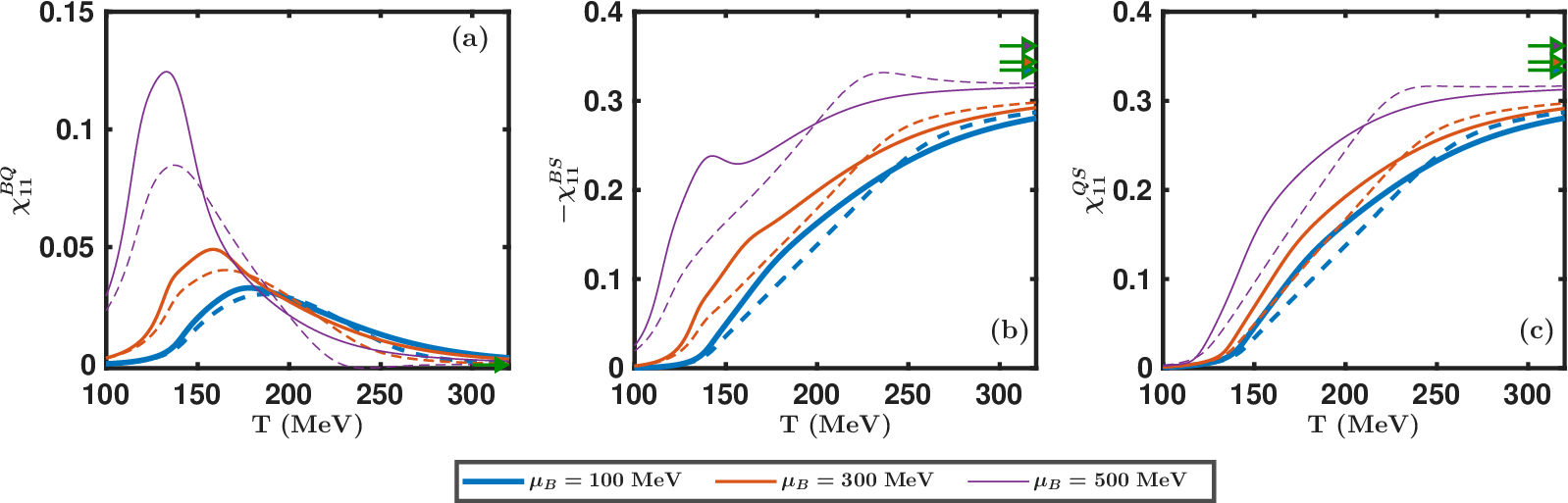}
		\caption{Second-order off-diagonal correlators (a) Baryon-charge $\chioff{11}{BQ}$, (b) baryon-strangeness $-\chioff{11}{BS}$, (c) charge-strangeness $\chioff{11}{QS}$ vs temperature at $\mub = 100$, 300, and 500 MeV. Each $\mub$ slice has its own color, and thicker lines mark lower $\mub$. Sign convention as in Fig. \ref{fig:offdiag2}. Solid (dashed) lines correspond to PCQMF with (without) the fermion vacuum term. Forest-green right-edge arrows mark the SB limits. In panels (b) and (c) the SB limit is $\mub$-dependent at finite $\mub$, so each slice carries its own arrow with the arrowhead filled by slice color (Appendix \ref{app:SB}).}
		\label{fig:offdiag_finmu}
	\end{figure*}
	
	Figure \ref{fig:offdiag_finmu} presents the second-order off-diagonal correlators $\chioff{11}{BQ}$, $-\chioff{11}{BS}$, and $\chioff{11}{QS}$ at $\mub = 100, 300, 500$ MeV. Each $\mub$ slice carries a distinct color, and line thickness decreases as $\mub$ grows. SB limits appear as forest-green arrows on the right edge of each panel.
	
	In panel (a), $\chioff{11}{BQ}$ peaks near the chiral crossover at every slice, and the peak position shifts to lower $T$ with increasing $\mub$. At $\mub = 500$ MeV, the vac=1 peak reaches $0.125$ at $T \approx 133$ MeV. The vac=0 curve follows the same pattern with smaller amplitudes. Light quarks carry both $B$ and $Q$, so the correlator couples directly to $\mub$ and the peak sharpens as the crossover strengthens. The vac=1 peak grows by a factor of $3.8$ between $\mub = 100$ and $500$ MeV, more than double the $\chiB{2}$ growth of $1.7$ (Fig. \ref{fig:chi24_finitemuB}).
	
	In panels (b) and (c), $-\chioff{11}{BS}$ and $\chioff{11}{QS}$ rise through the crossover toward the SB limit $\frac{1}{3} + \frac{\muhat^2}{9\pi^2}$. $-\chioff{11}{BS}$ develops increasing structure at larger $\mub$. At $\mub = 500$ MeV, the vac=0 curve shows a broad maximum near $T \approx 237$ MeV with a weaker structure at lower $T$. In vac=1, the low-$T$ structure near the chiral crossover dominates the curve. The chiral crossover affects the baryon sector more strongly, so the vac=1/vac=0 separation is larger in $\chioff{11}{BQ}$ than in $-\chioff{11}{BS}$. Unlike $\chioff{11}{BQ}$, which vanishes in the SB limit and grows as a pure crossover feature, $-\chioff{11}{BS}$ and $\chioff{11}{QS}$ approach a nonzero SB limit. The $\mub$ dependence therefore appears chiefly in the onset temperature and steepness of the rise. The vac=1 curves rise more steeply through the crossover (Sec. \ref{sec:order_params}), and both variants converge toward the SB limit at high $T$.
	
	\subsection{Fourth-order off-diagonal correlators at finite $\mub$}
	\label{sec:offdiag4_finite}
	
	\begin{figure*}[t]
		\centering
		\includegraphics[width=\textwidth]{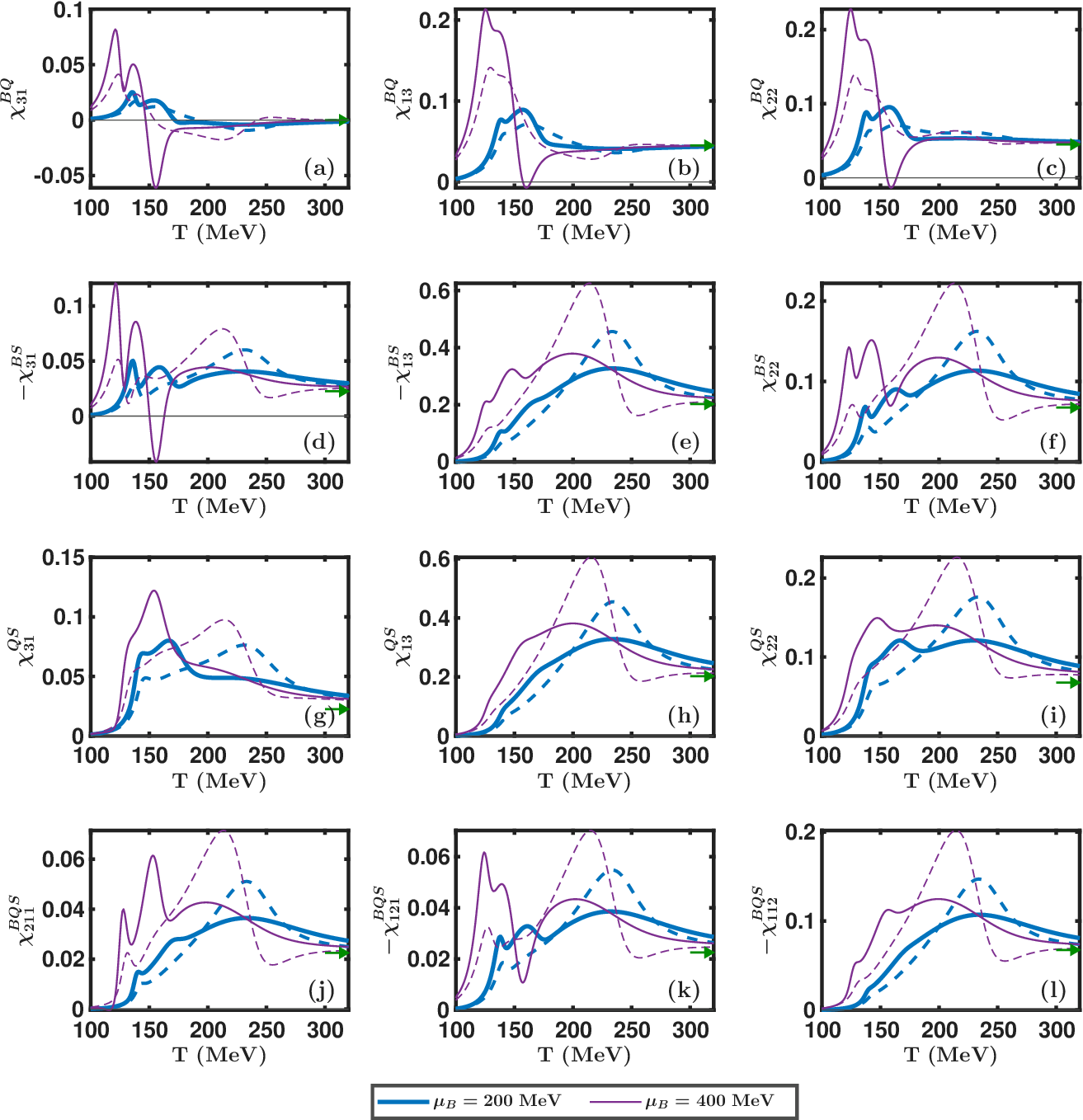}
		\caption{Complete fourth-order off-diagonal BQS tensor vs temperature at $\mub = 200$ and 400 MeV. Rows 1-3 show the two-charge correlators $\chioff{ij}{BQ,BS,QS}$ with $i+j=4$, columns ordered as $\chi_{31}$, $\chi_{13}$, $\chi_{22}$. Row 4 shows the mixed correlators $\chioff{ijk}{BQS}$ with $i+j+k=4$, columns ordered as $\chi_{211}$, $-\chi_{121}$, $-\chi_{112}$. Panels (d), (e), (k), (l) carry an overall minus sign to match the convention of Abuali et al.\ 2025 \cite{Abuali:2025WB}. Each $\mub$ slice has its own color, and thicker lines mark lower $\mub$. Solid (dashed) lines correspond to PCQMF with (without) the fermion vacuum term. Forest-green right-edge arrows mark the SB limits (Appendix \ref{app:SB}).}
		\label{fig:offdiag4_finmu}
	\end{figure*}
	
	Figure \ref{fig:offdiag4_finmu} presents the twelve independent fourth-order off-diagonal correlators at $\mub = 200$ and $400$ MeV. The panel ordering and sign convention match Fig. \ref{fig:offdiag4}. Each $\mub$ slice carries a distinct color, and line thickness decreases as $\mub$ grows. SB limits appear as forest-green arrows on the right edge. The same twelve correlators have been computed at $\mub = 0$ MeV in (2+1)-flavor PQM with the vacuum term \cite{Chatterjee:2012}, and in lattice QCD \cite{Abuali:2025WB}. A nuclear-matter calculation \cite{Yang:2025} reports the BQ subset at finite $\mub$ and finds $\chioff{31}{BQ} > \chioff{22}{BQ} > \chioff{13}{BQ}$ along the chemical freeze-out line. No prior calculation has reported the full twelve-component fourth-order off-diagonal tensor at finite $\mub$.
	
	Many vac=1 panels carry a double-peak fine structure in the chiral region at finite $\mub$ that resolves the chiral-deconfinement splitting. In panel (a), the vac=1 curve develops twin chiral peaks at both $\mub$ values. At $\mub = 200$ MeV, the peaks lie near $T = 135$ and $154$ MeV. At $\mub = 400$ MeV, they sharpen and shift to $T = 121$ and $136$ MeV, and the dip between them deepens. The vac=0 curve carries the same structure at a lower amplitude. The same double-peak structure appears in vac=1 at $\mub = 200$ MeV in panels (b), (c), (d), (f), (g), (j), and (k). It dominates the curve in panels (b), (c), and (g), and lies alongside a strange-melting peak in panels (d), (f), (j), and (k). At $\mub = 400$ MeV, the twin sharpens further in panels (a), (d), (f), (j), and (k). In panels (b), (c), and (g), the twin instead merges into a single tall, sharp peak. $\Tpc$ drops faster than $\Tdec$ with increasing $\mub$, and the chiral peak grows in amplitude. In panels (b), (c), and (g), the growing chiral peak absorbs the deconfinement peak. The vac=0 variant carries the same splitting in panels (a), (c), and (d) at $\mub = 200$ MeV with smaller amplitudes throughout. Larger chiral-peak amplitudes in vac=1 make the splitting more visible than in vac=0.
	
	At $\mub = 400$ MeV, the vac=1 curve dips below zero after the chiral peaks in panels (a), (b), (c), and (d). In panel (a), the vac=1 curve crosses zero after the chiral peaks and reaches a deep minimum at $T = 156$ MeV. The vac=1 curve in panel (d) shows the same deep dip at $T = 156$ MeV alongside the strange-melting peak. Panels (b) and (c) carry shallower troughs near $T \approx 160$ MeV after the tall chiral peak. The vac=0 curve dips below zero only in panel (a) and stays positive in panels (b), (c), and (d). The taller chiral peaks in vac=1 drive the curve below zero between the peaks. Beyond the chiral region, the curves carry additional structure. In most BS, QS, and BQS panels, a strange-melting peak dominates the curve at higher temperatures. The vac=0 strange peaks exceed the vac=1 strange peaks, the reverse of the chiral-region hierarchy. At $\mub = 400$ MeV in vac=1, the growing chiral peak overtakes the strange-melting peak in panels (d), (f), (j), and (k). The vac=1 chiral peaks exceed those of vac=0, while the vac=0 strange-melting peaks exceed those of vac=1.
	
	
	\subsection{Diagonal baryon susceptibilities at finite $\mub$}
	\label{sec:chiB_finite}
	
	\begin{figure*}[t]
		\centering
		\includegraphics[width=\textwidth]{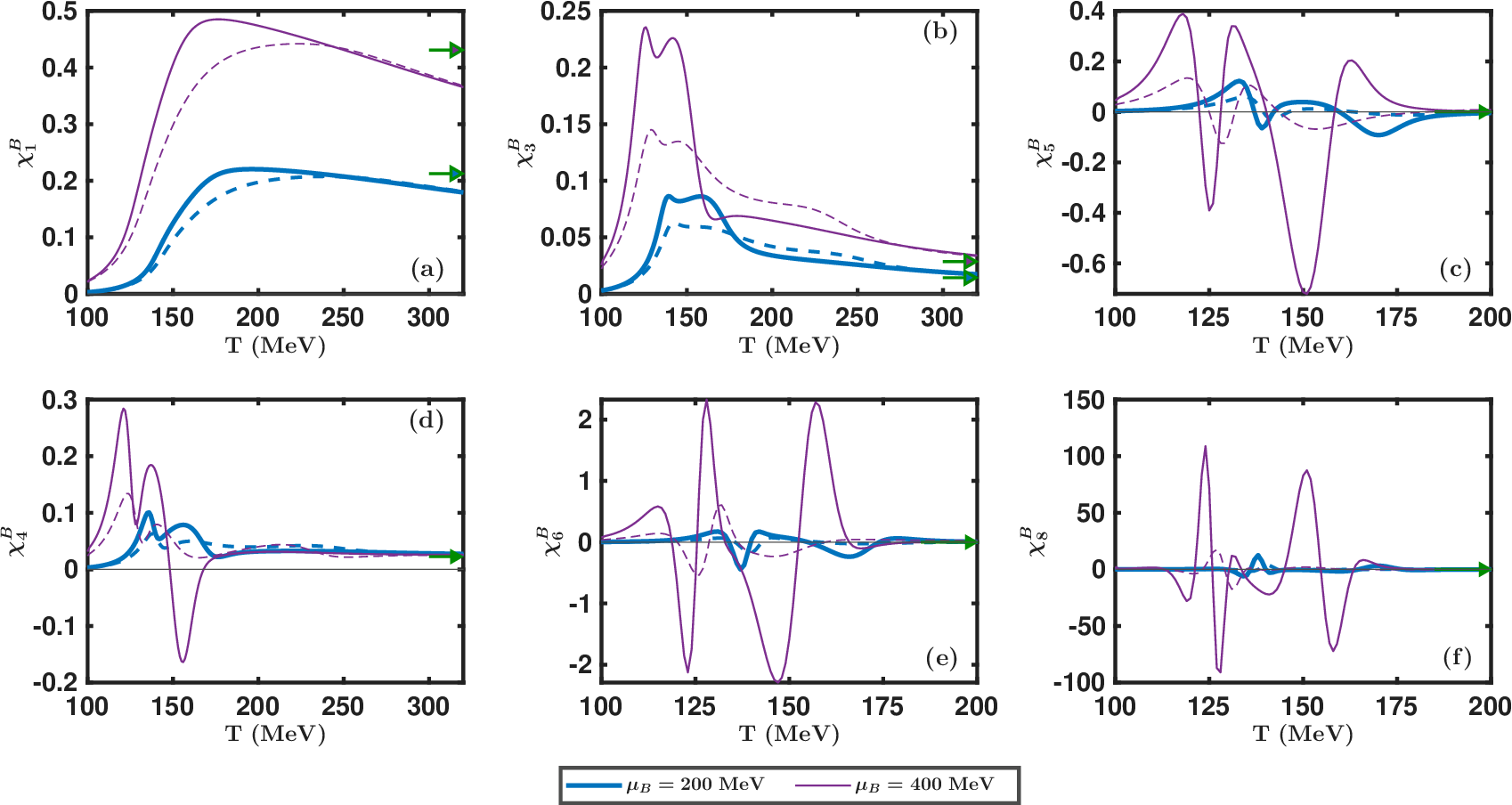}
		\caption{Diagonal baryon susceptibilities at $\mub = 200$ and 400 MeV. The top row shows odd-order (a) $\chiB{1}$, (b) $\chiB{3}$, and (c) $\chiB{5}$. The bottom row shows even-order (d) $\chiB{4}$, (e) $\chiB{6}$, and (f) $\chiB{8}$. Each $\mub$ slice has its own color, and thicker lines mark lower $\mub$. Solid (dashed) lines correspond to PCQMF with (without) the fermion vacuum term. Forest-green right-edge arrows mark the SB limits. For the slice-dependent SB values in panels (a) and (b), the arrowhead is filled by slice color (Appendix \ref{app:SB}).}
		\label{fig:chi_odd}
	\end{figure*}
	
	Figure \ref{fig:chi_odd} presents the diagonal baryon susceptibilities $\chiB{n}$ at $\mub = 200$ and $400$ MeV. The top row shows the odd-order $\chiB{1}$, $\chiB{3}$, $\chiB{5}$ and the bottom row the even-order $\chiB{4}$, $\chiB{6}$, $\chiB{8}$. The odd-order $\chiB{n}$ vanish at $\mub = 0$ MeV by charge-conjugation symmetry and are accessible only at finite $\mub$. Each $\mub$ slice carries a distinct color, and line thickness decreases as $\mub$ grows. SB limits appear as right-edge arrows. Panels (a) and (b) carry two arrows each because $\chiB{1}|_\SB$ and $\chiB{3}|_\SB$ depend on $\mub/T$ and take distinct values at each slice. The remaining panels carry a single $\mub$-independent SB arrow.
	
	The odd-order $\chiB{n}$ develops richer chiral-region structure as $n$ grows. In Fig. \ref{fig:chi_odd}(a), $\chiB{1}$ rises monotonically through the crossover and saturates at high $T$ at both $\mub$ values. The vac=1 curve lies above vac=0 in the chiral crossover region, and the two merge at high $T$. In Fig. \ref{fig:chi_odd}(b), $\chiB{3}$ remains positive across the full $T$ range and carries twin peaks in vac=1 at both $\mub$ values. The peaks shift to lower $T$ and sharpen as $\mub$ grows. The vac=0 curve shows only a single broad maximum at $\mub = 200$ MeV and develops the same twin structure at $\mub = 400$ MeV. A third broad bump develops at higher $T$ in both variants at $\mub = 400$ MeV. The twin peaks resolve the chiral-deconfinement splitting through the glue-to-Yang-Mills mapping \cite{Haas:2013}. In Fig. \ref{fig:chi_odd}(c), $\chiB{5}$ oscillates strongly across the chiral region in both variants. In vac=1 at $\mub = 400$ MeV, the curve develops a large negative minimum. The vac=0 extrema occur at smaller amplitudes and at higher temperatures than vac=1 at both $\mub$ values.
	
	The even-order $\chiB{n}$ develops sign reversals and large oscillation amplitudes across the chiral region. In Fig. \ref{fig:chi_odd}(d), $\chiB{4}$ repeats the double peak and sign reversal of Fig. \ref{fig:chi24_finitemuB}(d) at these slices. The sign reversal in $\chiB{4}$ and the off-diagonal sign reversals in Fig. \ref{fig:offdiag4_finmu} share the same mechanism. PNJL studies of fourth-order net-baryon cumulants near a critical endpoint report sign-changing $\chiB{4}$ along freeze-out lines \cite{Shao:2018,Ferreira:2018}. In those studies, proximity to the CEP drives the effect. In Fig. \ref{fig:chi_odd}(e), $\chiB{6}$ oscillates strongly with both signs across the chiral region. The vac=1 curve carries more visible extrema than vac=0 at both $\mub$ values. Amplitudes grow substantially from $\mub = 200$ to $\mub = 400$ MeV. In Fig. \ref{fig:chi_odd}(f), $\chiB{8}$ shows the most violent oscillation of the figure. At $\mub = 400$ MeV, the vac=1 curve oscillates with multiple sign changes across the narrow $T$ range. The vac=1 amplitudes far exceed those of vac=0. The fRG calculation of hyperorder net-baryon fluctuations through tenth order at finite $\mub$ \cite{Fu:2021_hyperorder} reports a similar growth of chiral-region oscillations with $\mub$. PQM with and without the vacuum term reports the corresponding $R_{n2}^B$ ratios through twelfth order at finite $\mub$ \cite{Schaefer:2012} but does not isolate individual high-order $\chiB{n}$.

	\subsection{Cumulant ratios along the pseudocritical line}
	\label{sec:ratios_Tpc}
	
	\begin{figure*}[t]
		\centering
		\includegraphics[width=0.80\textwidth]{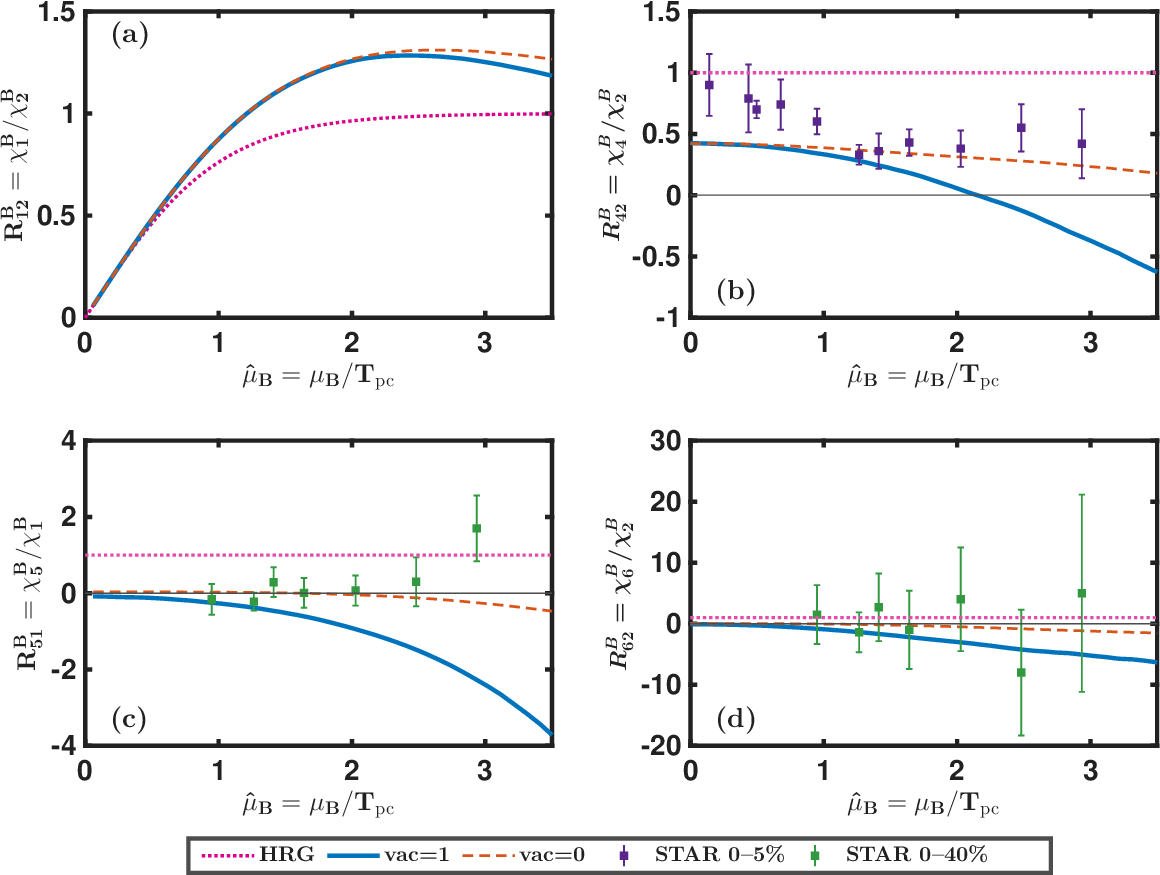}
		\caption{Cumulant ratios along the pseudocritical line $\Tpc(\mub)$ as functions of $\muhat = \mub/\Tpc$. (a) $R_{12}^B = \chiB{1}/\chiB{2}$ with HRG reference line $\tanh\muhat$ (magenta dotted), (b) $R_{42}^B = \chiB{4}/\chiB{2}$, (c) $R_{51}^B = \chiB{5}/\chiB{1}$, and (d) $R_{62}^B = \chiB{6}/\chiB{2}$. Solid (dashed) lines correspond to PCQMF with (without) the fermion vacuum term. STAR BES-II net-proton data are overlaid in panel (b) (0-5\% centrality \cite{STAR:2025_BES2}, purple squares) and in panels (c) and (d) (0-40\% centrality \cite{STAR:2025_C5C6}, green squares). Magenta dotted horizontal line in panels (b)-(d) marks the HRG baseline $R_{n2}^B = 1$.}
		\label{fig:ratios_Tpc}
	\end{figure*}
	
	Figure \ref{fig:ratios_Tpc} presents the cumulant ratios $R_{12}^B \equiv \chiB{1}/\chiB{2}$, $R_{42}^B \equiv \chiB{4}/\chiB{2}$, $R_{51}^B \equiv \chiB{5}/\chiB{1}$, and $R_{62}^B \equiv \chiB{6}/\chiB{2}$ along the pseudocritical line $\Tpc(\mub)$ as functions of $\muhat = \mub/\Tpc$. The HRG baseline $\tanh\muhat$ is shown in panel (a) and $R_{n2}^B = 1$ in panels (b), (c), and (d) \cite{Friman:2011,Karsch:2011}. STAR BES-II net-proton data are overlaid in panels (b), (c), and (d) \cite{STAR:2025_BES2,STAR:2025_C5C6}. The STAR points lie on the chemical freeze-out trajectory under the heavy-ion constraints of strangeness neutrality, $\langle n_S \rangle = 0$, and fixed electric-to-baryon ratio, $n_Q/n_B = 0.4$. The model curves are evaluated at $\Tpc(\mub)$ with $\muq = \mus = 0$. A comparison that imposes the heavy-ion constraints on the model is left for future work.
	
	The ratios $R_{12}^B$ and $R_{42}^B$ depart from the HRG baselines as $\muhat$ grows, and vac=1 develops a sign reversal at fourth order. In Fig. \ref{fig:ratios_Tpc}(a), $R_{12}^B$ rises monotonically from zero to a maximum near $\muhat = 2.5$ in both variants. The vac=1 and vac=0 curves overlap across most of the range and separate only at high $\muhat$. Both substantially exceed the HRG reference $\tanh\muhat$ above $\muhat \approx 0.5$. The HRG limit no longer applies because mean-field quark degrees of freedom dominate at $\Tpc(\mub)$. In Fig. \ref{fig:ratios_Tpc}(b), both variants start at $R_{42}^B \approx 0.42$ at $\muhat = 0$. The vac=1 curve decreases monotonically and crosses zero at $\muhat \approx 2.15$. The curve continues to large negative values at higher $\muhat$. The vac=0 curve also decreases but stays positive across the full $\muhat$ range. In vac=1, $\chiB{4}$ develops the sign reversal seen in Fig. \ref{fig:chi_odd}(d), and $R_{42}^B$ inherits the sign change because $\chiB{2}$ is positive throughout. STAR central values lie above both model curves at low $\muhat$ and remain positive throughout.
	
	In vac=1, $R_{51}^B$ and $R_{62}^B$ start negative and grow more negative as $\muhat$ increases. In vac=0, both ratios start marginally positive and become negative only at higher $\muhat$. In Fig. \ref{fig:ratios_Tpc}(c), the vac=0 curve becomes negative at $\muhat \approx 1.6$. The vac=1 curve falls within the STAR error bars at the two lowest-$\muhat$ points and below them at higher $\muhat$, while the vac=0 curve falls within the bars except at the largest $\muhat$. In Fig. \ref{fig:ratios_Tpc}(d), the vac=0 curve becomes negative earlier at $\muhat \approx 0.76$. Both variants fall within the STAR error bars throughout, since the experimental uncertainties at sixth order are large. The vac=1 amplitudes far exceed vac=0 in both panels and grow faster with $\muhat$ at sixth order. The vac=1 amplification follows the same pattern that drives the high-order $\chiB{n}$ amplification in Fig. \ref{fig:chi_odd}(e) and (f).
	
	\subsection{Sensitivity to the vector coupling}
	\label{sec:gv_sensitivity}
	
\begin{figure*}[t]
		\centering
		\includegraphics[width=\textwidth]{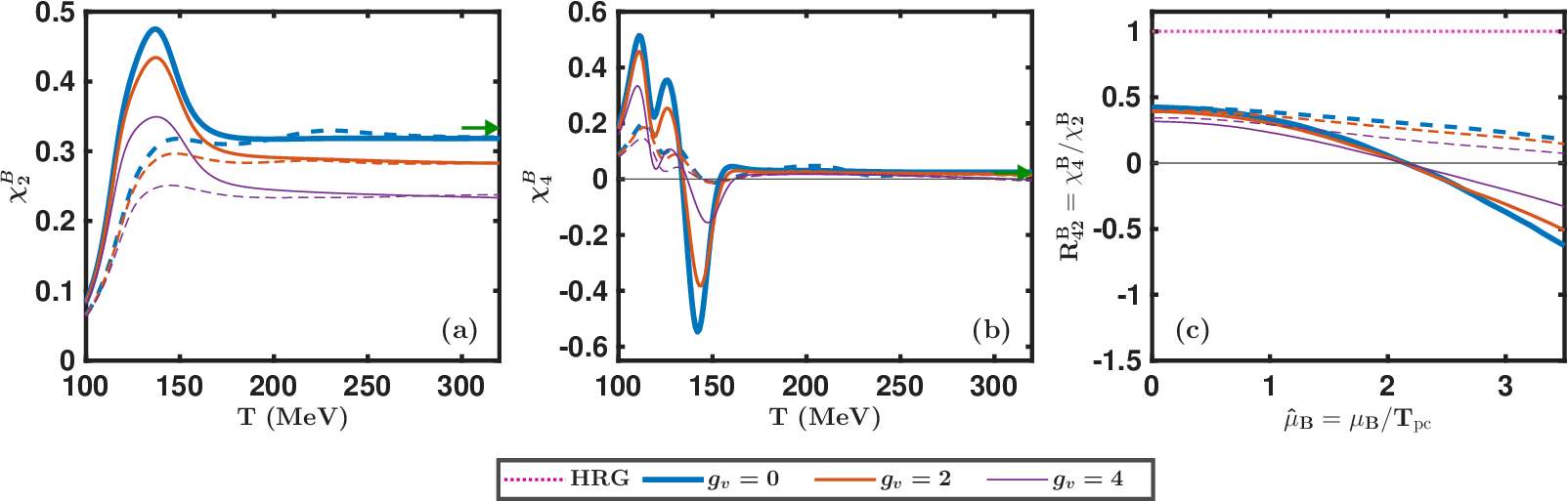}
		\caption{Diagonal baryon susceptibilities and their fourth-to-second cumulant ratio for $g_v = 0$, 2, and 4. (a) $\chiB{2}$ and (b) $\chiB{4}$ vs temperature at $\mub = 500$ MeV. (c) $R_{42}^B = \chiB{4}/\chiB{2}$ along the pseudocritical line $\Tpc(\mub)$ as a function of $\muhat = \mub/\Tpc$. Each $g_v$ value has its own color, and thicker lines mark lower $g_v$. Solid (dashed) lines correspond to PCQMF with (without) the fermion vacuum term. Forest-green right-edge arrows in panels (a) and (b) mark the SB limits (Appendix \ref{app:SB}). Magenta dotted horizontal line in panel (c) marks the HRG baseline $R_{42}^B = 1$.}
		\label{fig:gv_sensitivity}
	\end{figure*}

	The vector coupling is held at $g_v = 0$ in the main results. Its effect on the finite-$\mub$ baryon sector is assessed here. Figure \ref{fig:gv_sensitivity} presents $\chiB{2}$ and $\chiB{4}$ versus temperature at $\mub = 500$ MeV in panels (a) and (b), and the ratio $R_{42}^B = \chiB{4}/\chiB{2}$ along the pseudocritical line $\Tpc(\mub)$ as a function of $\muhat = \mub/\Tpc$ in panel (c). A nonzero $g_v$ generates repulsive vector mean fields that lower the effective quark chemical potentials and weaken the baryon-number response. In panel (a), $\chiB{2}$ at $\mub = 500$ MeV declines monotonically with $g_v$, and the crossover peak drops as $g_v$ rises from 0 to 4. The fourth-order response in panel (b) retains the two maxima below the crossover and the negative dip of Fig. \ref{fig:chi24_finitemuB}(d) and Fig. \ref{fig:chi_odd}(d), but its amplitude falls with $g_v$. The vac=0 curves stay far flatter than the vac=1 curves throughout. The fourth-order oscillation is therefore a vacuum-term effect that $g_v$ only dampens.
	
	The features that distinguish the two model variants survive at every $g_v$. In panel (c), $R_{42}^B$ starts below the HRG baseline at $\mub = 0$ and decreases with $\muhat$. The vac=1 curves cross zero and turn negative, while the vac=0 curves remain positive over the full range, as in Fig. \ref{fig:ratios_Tpc}(b). The zero crossing sits at $\muhat \approx 2.15$ regardless of $g_v$. A larger $g_v$ suppresses the susceptibilities without altering their sign structure. The choice $g_v = 0$ in the main results, therefore, leaves every qualitative conclusion of this work unchanged.
	
	\section{Discussion}
	\label{sec:discussion}
	
	The fermion vacuum term of Eq. (\ref{eq:Omega_vac}) enters the PCQMF effective potential and modifies the gap-equation solutions. In vac=1, Fig. \ref{fig:order_params}(a) shows that the scalar field $\sigma$ drops more steeply with temperature through the chiral crossover than in vac=0. This sharpening is opposite to the (2+1)-flavor PQM result, where the vacuum term smooths the crossover \cite{Chatterjee:2012,Schaefer:2012}. The light constituent quark masses follow, since they are set by the scalar fields through Eq. (\ref{eq:mstar}). In Fig. \ref{fig:phase_diagram}(b), $-d\Dls/dT$ develops an inflection near $\Tdec$ in vac=1 and rises monotonically into the chiral peak in vac=0. The deconfinement-driven inflection in $\Dls(T)$ near $\Tdec$ becomes visible in vac=1 through the coupled gap equations. The chiral-deconfinement splitting is generated by the glue-to-Yang-Mills mapping of Eq. (\ref{eq:TYM_Tglue}) \cite{Haas:2013} and is present in both variants. At higher derivative orders, Fig. \ref{fig:chi_higher_B} shows the chiral-deconfinement splitting resolved as a double-peak structure in $\chiB{4}$ and as multiple zero crossings in $\chiB{6}$ and $\chiB{8}$ in both variants. The vac=1 peak-to-trough contrast exceeds that of vac=0.
	
	The vac=1/vac=0 difference in the fourth-order off-diagonal tensor (Fig. \ref{fig:offdiag4}) depends on the temperature region of each peak. Chiral-crossover peaks are larger in vac=1, while strange-melting peaks are larger in vac=0. The BQ row sits in the first category, since light quarks dominate its correlators at the chiral crossover. The steeper $\sigma(T)$ drop in vac=1 drives the vac=1 dominance throughout the chiral region of the tensor, where the chiral-deconfinement double-peak resolves at low $T$. The dominant peaks of the BS, QS, and BQS rows sit in the second category, where vac=0 produces the higher amplitudes despite the steeper $\zeta(T)$ drop in vac=1 (Fig. \ref{fig:order_params}(b)). Both variants reproduce the same vacuum observables (Sec. \ref{sec:vacuum}), so the strange-melting reversal lies in the high-$T$ response of the two parameter sets rather than in the chiral-field response at low $T$ alone. 
	
	The qualitative differences between vac=1 and vac=0 persist at finite $\mub$. The chiral-region vac=1 dominance and the strange-region vac=0 dominance both persist into the finite-density region (Sec. \ref{sec:chi24_finite}-\ref{sec:offdiag4_finite}). Along $\Tpc(\mub)$, $R_{42}^B$ changes sign in vac=1 but stays positive in vac=0, and $R_{51}^B$ and $R_{62}^B$ are persistently negative in vac=1 (Sec. \ref{sec:ratios_Tpc}). 
	
	PCQMF treats the scalar fields ($\sigma$, $\zeta$, $\delta$) and the Polyakov loop ($\Ploop$, $\Ploopbar$) at the mean-field level, and fluctuations of these fields are absent. All three diagonal $\chi_2$ underestimate the lattice (Sec. \ref{sec:chi2}), and the deficit is largest for $\chiQ{2}$. Pions are the lightest charged carriers and dominate $\chiQ{2}$ in the hadronic phase, but in PCQMF they freeze at classical mean-field values and contribute no fluctuations. At low temperatures, PCQMF does not approach the HRG limit because hadrons are not realized as quasi-particle states. Beyond-mean-field methods such as the fRG incorporate both bosonic and fermionic fluctuations \cite{Fu:2021_hyperorder}.

	\section{Summary and outlook}
	\label{sec:summary}
	
	This work reports the first calculation of generalized susceptibilities of conserved charges in the PCQMF model with the fermion vacuum term. At $\mub = 0$ MeV, the calculation covers all diagonal $\chi_n^{B,Q,S}$ through eighth order and all twelve independent fourth-order off-diagonal correlators. An independently refitted no-sea variant (vac=0) sets the comparison baseline. The chiral pseudocritical temperature is $\Tpc = 170.5$ MeV (vac=1) and $166.4$ MeV (vac=0), and the deconfinement counterpart is $\Tdec = 144.4$ MeV and $146.6$ MeV. The chiral-deconfinement splitting reaches $26.1$ MeV in vac=1 and $19.8$ MeV in vac=0. The inflection in $-d\Dls/dT$ near $\Tdec$ appears in vac=1 only. Higher derivative orders resolve the chiral-deconfinement splitting as twin maxima in $\chiB{4}$ and $\chiQ{6}$, as twin minima in $\chiB{8}$ and $\chiQ{8}$, and as multiple zero crossings in $\chiB{6}$. Among the fourth-order off-diagonal correlators, the BQ components carry vac=1 dominance across the chiral crossover. The dominant peaks of the BS, QS, and mixed BQS components occur at higher strange-melting temperatures, where vac=0 dominates. The model reproduces the qualitative crossover behavior seen on the lattice. Channel-by-channel magnitude differences are detailed in Sec. \ref{sec:results_mub0}. 
	
	The calculation extends to finite baryon chemical potential up to $\mub = 500$ MeV at $\muq = \mus = 0$. Along the pseudocritical line $\Tpc(\mub)$, $R_{42}^B$ starts at $\approx 0.42$ in both variants and decreases monotonically with $\muhat = \mub/\Tpc$. The vac=1 curve crosses zero at $\muhat \approx 2.15$ and continues to large negative values, while vac=0 stays positive across the full range. $R_{51}^B$ and $R_{62}^B$ start negative in vac=1 and grow more negative as $\muhat$ increases. In vac=0, they start marginally positive and turn negative at $\muhat \approx 1.6$ and $\muhat \approx 0.76$ respectively. The vac=1 amplitudes far exceed vac=0 in both. The chiral and strange-melting dominance pattern observed at $\mub = 0$ persists into the finite-density region. The vac=1 curves dominate the chiral region, and the vac=0 curves dominate the strange-melting region. 
	
	In our calculations, we use $\muq = \mus = 0$ throughout, while STAR BES-II data lie on the heavy-ion freeze-out trajectory of strangeness neutrality and $n_Q/n_B = 0.4$ (Sec. \ref{sec:ratios_Tpc}). A direct quantitative comparison requires model evaluation on this trajectory and is left for future work. Imposing these constraints self-consistently determines $\muq(T,\mub)$ and $\mus(T,\mub)$ and enables direct comparison with the STAR cumulant ratios. Isentropes of constant entropy per net baryon density $s/n_B$ trace the hydrodynamic cooling path at fixed $\sqrt{s_{NN}}$ and connect model predictions to specific BES collision energies. Critical endpoint localization on the extended $(T, \mub)$ plane in both variants quantifies the vac=1/vac=0 shift. The earlier PCQMF endpoint estimate \cite{Kumari:2021} used neither the vacuum term nor the glue-to-Yang-Mills mapping.
	
	\begin{acknowledgments}
		DS sincerely acknowledges the support for this work from the Ministry of Science and Human Resources (MHRD), Government of India, through an Institute fellowship at the National Institute of Technology Jalandhar. AK sincerely	acknowledges the Anusandhan National Research Foundation (ANRF), Government of India, for funding the research project under the Science and Engineering
		Research Board-Core Research Grant (SERB-CRG) scheme (File	No. CRG/2023/000557). AK also sincerely acknowledges the DST-SERB, Government of India for funding the research project CRG/2019/000096 under which work on fluctuations of conserved charges initiated.
	\end{acknowledgments}

	\appendix
	\section{Stefan-Boltzmann limits}
	\label{app:SB}
	\renewcommand{\theequation}{A\arabic{equation}}
	\setcounter{equation}{0}
	
	In the high-temperature limit, the constituent quark masses approach zero, and the gluonic sector decouples, so the pressure reduces to that of a free three-flavor massless quark gas. The single-flavor pressure for one massless Dirac quark with $\Nc$ colors at $\hat\mu_f \equiv \mu_f/T$ reads \cite{Kapusta:2006}
	\begin{equation}
		\frac{p_f^{\SB}(T,\mu_f)}{T^4}
		= \Nc\!\left[\frac{7\pi^2}{180} + \frac{\hat\mu_f^2}{6} + \frac{\hat\mu_f^4}{12\pi^2}\right],
		\label{eq:p_SB_flavor}
	\end{equation}
	which is a quartic polynomial in $\hat\mu_f$. Differentiating with respect to $\hat\mu_f$ at $\Nc = 3$ gives the only nonzero single-flavor susceptibilities,
	\begin{align}
		\chi_1^{ff}\big|_\SB &= \hat\mu_f + \frac{\hat\mu_f^3}{\pi^2}, &
		\chi_2^{ff}\big|_\SB &= 1 + \frac{3\hat\mu_f^2}{\pi^2}, \nonumber \\
		\chi_3^{ff}\big|_\SB &= \frac{6\hat\mu_f}{\pi^2}, &
		\chi_4^{ff}\big|_\SB &= \frac{6}{\pi^2},
		\label{eq:chi_SB_flavor}
	\end{align}
	with $\chi_n^{ff}|_\SB = 0$ for $n \geq 5$. Generalized BQS susceptibilities follow from the linear charge-matrix relation between $\hat\mu_f$ and the conserved-charge potentials of Eq. (\ref{eq:mu_flavor}),
	\begin{equation}
		\chi_{ijk}^{BQS}\big|_\SB
		= \sum_{f=u,d,s} B_f^i\, Q_f^j\, S_f^k\, \chi_n^{ff}\big|_\SB,
		\qquad n = i+j+k,
		\label{eq:SB_master}
	\end{equation}
	with $B_f = 1/3$ for all flavors, $(Q_u, Q_d, Q_s) = (2/3, -1/3, -1/3)$, and $(S_u, S_d, S_s) = (0, 0, -1)$. The values at $\hat\mu_B = \hat\mu_Q = \hat\mu_S = 0$ for every quantity computed in this work are collected in Table \ref{tab:SB_limits}. All sixth- and eighth-order diagonal susceptibilities and all odd-order baryon susceptibilities with $n \geq 5$ vanish at SB, since $p_f^{\SB}/T^4$ is exactly quartic in $\hat\mu_f$.
	
	\begin{table*}[!htbp]
		\caption{Stefan-Boltzmann limits at $\hat\mu_B = \hat\mu_Q = \hat\mu_S = 0$ for the free three-flavor massless quark gas with $\Nc = 3$, covering every quantity computed in this work. Slice-dependent values at finite $\mub$ for $\chiB{1}$, $\chiB{2}$, $\chiB{3}$, $\chiQ{2}$, and $\chiS{2}$ are given in Eqs. (\ref{eq:SB_chi1B})-(\ref{eq:SB_chi2S}).}
		\label{tab:SB_limits}
		\begin{ruledtabular}
			\begin{tabular}{lclc}
				Observable & Expression &
				Observable & Expression \\
				\hline
				$\chiB{2}$ & $1/3$ &
				$\chioff{11}{BQ}$ & $0$ \\
				$\chiQ{2}$ & $2/3$ &
				$\chioff{11}{BS}$ & $-1/3$ \\
				$\chiS{2}$ & $1$ &
				$\chioff{11}{QS}$ & $1/3$ \\
				\hline
				$\chiB{4}$ & $2/(9\pi^2)$ &
				$\chioff{31}{BQ}$ & $0$ \\
				$\chiQ{4}$ & $4/(3\pi^2)$ &
				$\chioff{22}{BQ}$ & $4/(9\pi^2)$ \\
				$\chiS{4}$ & $6/\pi^2$ &
				$\chioff{13}{BQ}$ & $4/(9\pi^2)$ \\
				\hline
				$\chioff{31}{BS}$ & $-2/(9\pi^2)$ &
				$\chioff{31}{QS}$ & $2/(9\pi^2)$ \\
				$\chioff{22}{BS}$ & $2/(3\pi^2)$ &
				$\chioff{22}{QS}$ & $2/(3\pi^2)$ \\
				$\chioff{13}{BS}$ & $-2/\pi^2$ &
				$\chioff{13}{QS}$ & $2/\pi^2$ \\
				\hline
				$\chioff{211}{BQS}$ & $2/(9\pi^2)$ &
				$C_{BS}$ & $1$ \\
				$\chioff{121}{BQS}$ & $-2/(9\pi^2)$ &
				$\chioff{11}{BQ}/\chiB{2}$ & $0$ \\
				$\chioff{112}{BQS}$ & $-2/(3\pi^2)$ &
				$\chioff{11}{QS}/\chiS{2}$ & $1/3$ \\
				\hline
				$R_{42}^B$ & $2/(3\pi^2)$ &
				$\chiB{2}/\chiQ{2}$ & $1/2$ \\
				$R_{42}^Q$ & $2/\pi^2$ &
				$\chiB{5}, \chiB{6}, \chiQ{6}, \chiS{6}, \chiB{8}, \chiQ{8}, \chiS{8}$ & $0$ \\
				$R_{42}^S$ & $6/\pi^2$ &
				$R_{62}^B, R_{62}^Q, R_{62}^S, R_{82}^B$ & $0$ \\
			\end{tabular}
		\end{ruledtabular}
	\end{table*}
	
	Setting $\muq = \mus = 0$ in Eq. (\ref{eq:SB_master}) gives the $\muhat$-dependent SB limits
	\begin{equation}
		\chiB{1}\big|_\SB = \frac{\muhat}{3} + \frac{\muhat^3}{27\pi^2},
		\label{eq:SB_chi1B}
	\end{equation}
	\begin{equation}
		\chiB{2}\big|_\SB = \frac{1}{3} + \frac{\muhat^2}{9\pi^2},
		\label{eq:SB_chi2B}
	\end{equation}
	\begin{equation}
		\chiB{3}\big|_\SB = \frac{2\muhat}{9\pi^2}.
		\label{eq:SB_chi3B}
	\end{equation}
	The charge and strangeness second-order limits are
	\begin{equation}
		\chiQ{2}\big|_\SB = \frac{2}{3} + \frac{2\muhat^2}{9\pi^2},
		\label{eq:SB_chi2Q}
	\end{equation}
	\begin{equation}
		\chiS{2}\big|_\SB = 1 + \frac{\muhat^2}{3\pi^2}.
		\label{eq:SB_chi2S}
	\end{equation}
	$\chiB{4}|_\SB = 2/(9\pi^2)$ becomes independent of $\muhat$, and $\chiB{n}|_\SB = 0$ for $n \geq 5$. The slice-dependent SB arrows in Fig. \ref{fig:chi_odd}(a) and (b) evaluate Eqs. (\ref{eq:SB_chi1B}) and (\ref{eq:SB_chi3B}) at the arrow-head temperature $T = 315$ MeV. At $\mub = 200$ MeV this gives $\chiB{1}|_\SB = 0.2126$ and $\chiB{3}|_\SB = 0.01430$. At $\mub = 400$ MeV, $\chiB{1}|_\SB = 0.4310$ and $\chiB{3}|_\SB = 0.02860$. The slice-dependent SB arrows in the top row of Fig. \ref{fig:chi24_finitemuB} likewise evaluate Eqs. (\ref{eq:SB_chi2B}), (\ref{eq:SB_chi2Q}), and (\ref{eq:SB_chi2S}) at $T = 315$ MeV. At $\mub = 500$ MeV this gives $\chiB{2}|_\SB = 0.362$, $\chiQ{2}|_\SB = 0.723$, and $\chiS{2}|_\SB = 1.085$. The $\mub = 100$ and $300$ MeV arrows sit close to the $\muhat = 0$ values $1/3$, $2/3$, and $1$. Panels (b) and (c) of Fig. \ref{fig:offdiag_finmu} use the same slice-dependent form, $-\chioff{11}{BS}|_\SB = \chioff{11}{QS}|_\SB = 1/3 + \muhat^2/(9\pi^2)$, while panel (a) keeps $\chioff{11}{BQ}|_\SB = 0$.
	
	\bibliography{references}
	
\end{document}